\documentstyle[psfig,preprint,aps,eqsecnum,tighten,floats]{revtex}
\begin{document}
\draft
\title{Quantum transport in\\
semiconductor--superconductor microjunctions}
\author{C. W. J. Beenakker}
\address{Instituut-Lorentz, University of Leiden\\
P.O. Box 9506, 2300 RA Leiden, The Netherlands}
\date{June, 1994}
\maketitle
\begin{abstract}
Recent experiments on conduction between a semiconductor and a
superconductor have revealed a variety of new mesoscopic phenomena.
Here is a review of the present status of this rapidly developing
field. A scattering theory is described which leads to a conductance
formula analogous to Landauer's formula in normal-state conduction. The
theory is used to identify features in the conductance which can serve
as ``signatures'' of phase-coherent Andreev reflection, i.e.\ for which
the phase coherence of the electrons and the Andreev-reflected holes is
essential. The applications of the theory include a quantum point
contact (conductance quantization in multiples of $4e^2/h$), a quantum
dot (non-Lorentzian conductance resonance), and quantum interference
effects in a disordered normal-superconductor junction (enhanced
weak-localization and reflectionless tunneling through a potential
barrier). The final two sections deal with the effects of Andreev
reflection on universal conductance fluctuations and on the shot noise.\\
{\tt Lectures at the Les Houches summer school,
Session LXI, 1994, to be published in:}
{\em Mesoscopic Quantum Physics},
E. Akkermans, G. Montambaux, and J.-L. Pichard, eds.
(North-Holland, Amsterdam).
\end{abstract}
\newpage
\narrowtext

\section{Introduction}

\begin{figure}[tb]
\hspace*{\fill}
\psfig{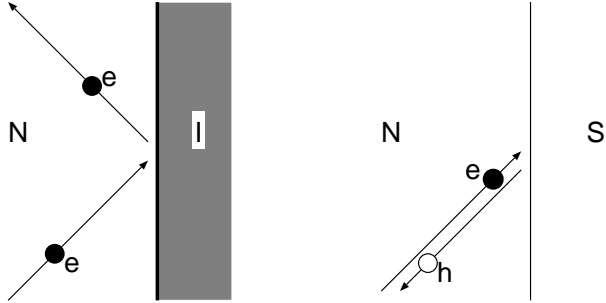}
\hspace*{\fill}
\medskip\caption[]{Normal reflection by an insulator (I) versus Andreev reflection by a
superconductor (S) of an electron excitation in a normal metal (N) near the
Fermi level. Normal reflection (left) conserves charge but does not conserve
momentum. Andreev reflection (right) conserves momentum but does not conserve
charge: The electron (e) is reflected as a hole (h) with the same momentum and
opposite velocity. The missing charge of $2e$ is absorbed as a Cooper pair by
the superconducting condensate.
\label{reflection}}
\end{figure}

At the interface between a normal metal and a superconductor, dissipative
electrical current is converted into dissipationless supercurrent. The
mechanism for this conversion was discovered thirty years ago by A.~F.~Andreev
\cite{And64}: An electron excitation slightly above the Fermi level in the
normal metal is reflected at the interface as a hole excitation slightly below
the Fermi level (see fig.\ \ref{reflection}). The missing charge of $2e$ is
removed as a supercurrent. The reflected hole has (approximately) the same
momentum as the incident electron. (The two momenta are precisely equal at the
Fermi level.) The velocity of the hole is minus the velocity of the electron
(cf.\ the notion of a hole as a ``time-reversed'' electron). This curious
scattering process is known as retro-reflection or {\em Andreev reflection}.

The early theoretical work on the conductance of a normal-metal --
superconductor (NS) junction treats the dynamics of the quasiparticle
excitations {\em semiclassically}, as is appropriate for macroscopic junctions.
Phase coherence of the electrons and the Andreev-reflected holes is ignored.
Interest in ``mesoscopic'' NS junctions, where phase coherence plays an
important role, is a recent development. Significant advances have been made
during the last few years in our understanding of quantum interference effects
due to phase-coherent Andreev reflection. Much of the motivation has come from
the technological advances in the fabrication of a highly transparent contact
between a superconducting film and the two-dimensional electron gas in a
semiconductor heterostructure. These systems are ideal for the study of the
interplay of Andreev reflection and the mesoscopic effects known to occur in
semiconductor nanostructures \cite{Eer91}, because of the large Fermi
wavelength, large mean free path, and because of the possibility to confine the
carriers electrostatically by means of gate electrodes. In this series of
lectures we review the present status of this rapidly developing field of
research.

To appreciate the importance of phase coherence in NS junctions, consider the
resistance of a normal-metal wire (length $L$, mean free path $l$). This
resistance increases monotonically with $L$. Now attach the wire to a
superconductor via a tunnel barrier (transmission probability $\Gamma$). Then
the resistance has a {\em minimum\/} when $L\simeq l/\Gamma$. The minimum
disappears if the phase coherence between the electrons and holes is destroyed,
by increasing the voltage or by applying a magnetic field. The resistance
minimum is associated with the crossover from a $\Gamma^{-1}$ to a
$\Gamma^{-2}$ dependence on the barrier transparency. The $\Gamma^{-2}$
dependence is as expected for tunneling into a superconductor, being a
two-particle process. The $\Gamma^{-1}$ dependence is surprising. It is as if
the Andreev-reflected hole can tunnel through the barrier without reflections.
This socalled ``reflectionless tunneling'' requires relatively transparent NS
interfaces, with $\Gamma\gtrsim l/L$. Semiconductor --- superconductor
junctions are convenient, since the Schottky barrier at the interface is much
more transparent than a typical dielectric tunnel barrier. The technological
effort is directed towards making the interface as transparent as possible. A
nearly ideal NS interface ($\Gamma\simeq 1$) is required if one wishes to study
how Andreev reflection modifies the quantum interference effects in the normal
state. (For $\Gamma\ll 1$ these are obscured by the much larger
reflectionless-tunneling effect.) The modifications can be quite remarkable. We
discuss two examples.

The first is weak localization. In the normal state, weak localization can not
be detected in the current--voltage ($I$--$V$) characteristic, but requires
application of a magnetic field. The reason is that application of a voltage
(in contrast to a magnetic field) does not break time-reversal symmetry. In an
NS junction, however, weak localization can be detected in the $I$--$V$
characteristic, because application of a voltage destroys the phase coherence
between electrons and holes. The result is a small dip in $\partial I/\partial
V$ versus $V$ around $V=0$ for $\Gamma\simeq 1$. On reducing $\Gamma$, the dip
crosses over to a peak due to reflectionless tunneling. The peak is much larger
than the dip, but the widths are approximately the same.

The second example is universal conductance fluctuations. In the normal state,
the conductance fluctuates from sample to sample with a variance which is
independent of sample size or degree of disorder. This is one aspect of the
universality. The other aspect is that breaking of time-reversal symmetry (by a
magnetic field) reduces the variance by precisely a factor of two. In an NS
junction, the conductance fluctuations are also size and disorder independent.
However, application of a time-reversal-symmetry breaking magnetic field has no
effect on the magnitude.

These three phenomena, weak localization, reflectionless tunneling, and
universal conductance fluctuations, are discussed in sections 4, 5, and 6,
respectively. Sections 2 and 3 are devoted to a description of the theoretical
method and to a few illustrative applications. The method is a scattering
theory, which relates the conductance $G_{\rm NS}$ of the NS junction to the
$N\times N$ transmission matrix $t$ in the normal state ($N$ is the number of
transverse modes at the Fermi level). In the limit of zero temperature, zero
voltage, and zero magnetic field, the relationship is
\begin{equation}
G_{\rm NS}=\frac{4e^{2}}{h}\sum_{n=1}^{N}
\frac{T_{n}^{2}}{(2-T_{n})^{2}},\label{GNS1}
\end{equation}
where the transmission eigenvalue $T_{n}$ is an eigenvalue of the matrix
product $tt^{\dagger}$. The same numbers $T_{n}$ ($n=1,2,\ldots N$) determine
the conductance $G_{\rm N}$ in the normal state, according to the Landauer
formula
\begin{equation}
G_{\rm N}=\frac{2e^{2}}{h}\sum_{n=1}^{N}T_{n}.\label{GN1}
\end{equation}
The fact that the same transmission eigenvalues determine both $G_{\rm N}$ and
$G_{\rm NS}$ means that one can use the same (numerical and analytical)
techniques developed for quantum transport in the normal state. This is a
substantial technical and conceptual simplification.

The scattering theory can also be used for other transport properties, other
than the conductance, both in the normal and the superconducting state. An
example, discussed in section 7, is the shot noise due to the discreteness of
the carriers. A doubling of the ratio of shot-noise power to current can occur
in an NS junction, consistent with the notion of Cooper pair transport in the
superconductor.

We conclude in section 8.

We restrict ourselves in this review (with one exception) to two-terminal
geometries, with a single NS interface. Equation (\ref{GNS1}), as well as the
Landauer formula (\ref{GN1}), only describe the two-terminal conductance. More
complex multi-terminal geometries, involving several NS interfaces, have been
studied theoretically by Lambert and coworkers \cite{Lam93a,Lam93b}, and
experimentally by Petrashov et al.\ \cite{Pet93}. Since we focus on
phase-coherent effects, most of our discussion concerns the linear-response
regime of infinitesimal applied voltage. A recent review by Klapwijk contains a
more extensive coverage of the non-linear response at higher voltages
\cite{Kla94}. The scattering approach has also been applied to the Josephson
effect in SNS junctions \cite{Bee91}, resulting in a formula for the
supercurrent--phase relationship in terms of the transmission eigenvalues
$T_{n}$ in the normal state. We do not discuss the Josephson effect here, but
refer to ref.\ \cite{Shima} for a review of mesoscopic SNS junctions. Taken
together, ref.\ \cite{Shima} and the present work describe a unified approach
to mesoscopic superconductivity.

\section{Scattering theory}

\begin{figure}[tb]
\hspace*{\fill}
\psfig{figure=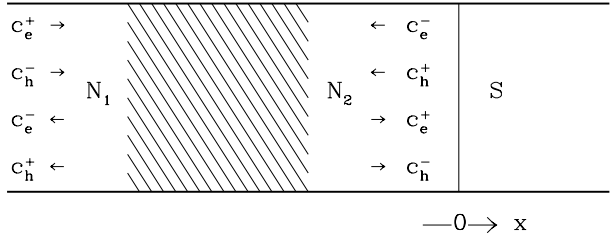,width=
8cm,bbllx=106pt,bblly=331pt,bburx=545pt,bbury=497pt}
\hspace*{\fill}
\medskip\caption[]{Normal-metal--superconductor junction containing a disordered normal
region (hatched). Scattering states in the two normal leads ${\rm N}_{1}$ and
${\rm N}_{2}$ are indicated schematically.
\label{diagram}}
\end{figure}

The model considered is illustrated in fig.\ \ref{diagram}. It consists of a
disordered normal region (hatched) adjacent to a superconductor (S). The
disordered region may also contain a geometrical constriction or a tunnel
barrier. To obtain a well-defined scattering problem we insert ideal
(impurity-free) normal leads ${\rm N}_{1}$ and ${\rm N}_{2}$ to the left and
right of the disordered region. The NS interface is located at $x=0$. We assume
that the only scattering in the superconductor consists of Andreev reflection
at the NS interface, i.e.\ we consider the case that the disorder is contained
entirely within the normal region. The spatial separation of Andreev and normal
scattering is the key simplification which allows us to relate the conductance
directly to the normal-state scattering matrix. The model is directly
applicable to a superconductor in the clean limit (mean free path in S large
compared to the superconducting coherence length $\xi$), or to a point-contact
junction (formed by a constriction which is narrow compared to $\xi$). In both
cases the contribution of scattering within the superconductor to the junction
resistance can be neglected \cite{Pip71}.

The scattering states at energy $\varepsilon$ are eigenfunctions of the
Bogoliubov-de Gennes (BdG) equation. This equation has the form of
two Schr\"{o}dinger equations for electron and hole wavefunctions ${\rm
u}(\vec{r})$ and ${\rm v}(\vec{r})$, coupled by the pair potential
${\mit\Delta}(\vec{r})$ \cite{deG66}:
\begin{eqnarray}
\left(\begin{array}{cc} {\cal H}_{0}&{\mit\Delta}\\ {\mit\Delta}^{\ast}&-{\cal
H}_{0}^{\ast} \end{array}\right) \left(\begin{array}{c}{\rm u}\\{\rm
v}\end{array}\right)=\varepsilon \left(\begin{array}{c}{\rm u}\\{\rm
v}\end{array}\right) .\label{BdG1}
\end{eqnarray}
Here ${\cal H}_{0}=(\vec{p}+e\vec{A})^{2}/2m+V-E_{\rm F}$ is the
single-electron Hamiltonian, containing an electrostatic potential $V(\vec{r})$
and vector potential $\vec{A}(\vec{r})$. The excitation energy $\varepsilon$ is
measured relative to the Fermi energy $E_{\rm F}$. To simplify construction of
the scattering basis we assume that the magnetic field $\vec{B}$ (in the
$z$-direction) vanishes outside the disordered region. One can then choose a
gauge such that $\vec{A}\equiv 0$ in lead ${\rm N}_{2}$ and in S, while
$A_{x},A_{z}=0$, $A_{y}=A_{1}\equiv{\rm constant}$ in lead ${\rm N}_{1}$.

The pair potential in the bulk of the superconductor ($x\gg\xi$) has amplitude
${\mit\Delta_{0}}$ and phase $\phi$. The spatial dependence of
${\mit\Delta}(\vec{r})$ near the NS interface is determined by the
self-consistency relation \cite{deG66}
\begin{equation}
{\mit\Delta}(\vec{r})=|g(\vec{r})|\sum_{\varepsilon>0}{\rm
v}^{\ast}(\vec{r}){\rm u}(\vec{r})[1-2f(\varepsilon)],\label{selfconsist}
\end{equation}
where the sum is over all states with positive eigenvalue, and
$f(\varepsilon)=[1+\exp(\varepsilon/k_{\rm B}T)]^{-1}$ is the Fermi function.
The coefficient $g$ is the interaction constant of the BCS theory of
superconductivity. At an NS interface, $g$ drops abruptly (over atomic
distances) to zero, in the assumed absence of any pairing interaction in the
normal region. Therefore, ${\mit\Delta}(\vec{r})\equiv 0$ for $x<0$. At the
superconducting side of the NS interface, ${\mit\Delta}(\vec{r})$ recovers its
bulk value $\Delta_{0}{\rm e}^{{\rm i}\phi}$ only at some distance from the
interface. We will neglect the suppression of ${\mit\Delta}(\vec{r})$ on
approaching the NS interface, and use the step-function model
\begin{equation}
{\mit\Delta}(\vec{r})={\mit\Delta_{0}}{\rm e}^{{\rm
i}\phi}\theta(x).\label{stepfunction}
\end{equation}
This model is also referred to in the literature as a ``rigid
boundary-condition''. Likharev \cite{Lik79} discusses in detail the conditions
for its validity: If the width $W$ of the NS junction is small compared to
$\xi$, the non-uniformities in ${\mit\Delta}(\vec{r})$ extend only over a
distance of order $W$ from the junction (because of ``geometrical dilution'' of
the influence of the narrow junction in the wide superconductor). Since
non-uniformities on length scales $\ll\xi$ do not affect the dynamics of the
quasiparticles, these can be neglected and the step-function model holds. A
point contact or microbridge belongs in general to this class of junctions.
Alternatively, the step-function model holds also for a wide junction if the
resistivity of the junction region is much bigger than the resistivity of the
bulk superconductor. This condition is formulated more precisely in ref.\
\cite{Lik79}. A semiconductor --- superconductor junction is typically in this
second category. Note that both the two cases are consistent with our
assumption that the disorder is contained entirely within the normal region.

It is worth emphasizing that the absence of a pairing interaction in the normal
region ($g(\vec{r})\equiv 0$ for $x<0$) implies a vanishing pair potential
${\mit\Delta}(\vec{r})$, according to eq.\ (\ref{selfconsist}), but does not
imply a vanishing order parameter $\Psi(\vec{r})$, which is given by
\begin{equation}
\Psi(\vec{r})=\sum_{\varepsilon>0}{\rm v}^{\ast}(\vec{r}){\rm
u}(\vec{r})[1-2f(\varepsilon)].\label{orderparam}
\end{equation}
Phase coherence between the electron and hole wave functions u and v leads to
$\Psi(\vec{r})\neq 0$ for $x<0$. The term ``proximity effect'' can therefore
mean two different things: One is the suppression of the pair potential
${\mit\Delta}$ at the superconducting side of the NS interface. This is a small
effect which is neglected in the present work (and in most other papers in this
field). The other is the induction of a non-zero order parameter $\Psi$ at the
normal side of the NS interface. This effect is fully included here, even
though $\Psi$ does not appear explicitly in the expressions which follow. The
reason is that the order parameter quantifies the degree of phase coherence
between electrons and holes, but does not itself affect the dynamics of the
quasiparticles. (The BdG equation (\ref{BdG1}) contains ${\mit\Delta}$ not
$\Psi$.)

We now construct a basis for the scattering matrix ($s$-matrix).
In the normal lead ${\rm N}_{2}$ the eigenfunctions of the BdG equation
(\ref{BdG1}) can be written in the form
\begin{eqnarray}
&&\psi_{n,{\rm e}}^{\pm}({\rm N}_{2})=
{\renewcommand{\arraystretch}{0.6}
\left(\begin{array}{c}1\\ 0\end{array}\right)}
(k_{n}^{\rm e})^{-1/2}\,\Phi_{n}(y,z)
\exp(\pm{\rm i}k_{n}^{\rm e}x),\nonumber\\
&&\psi_{n,{\rm h}}^{\pm}({\rm N}_{2})=
{\renewcommand{\arraystretch}{0.6}
\left(\begin{array}{c}0\\ 1\end{array}\right)}
(k_{n}^{\rm h})^{-1/2}\,\Phi_{n}(y,z)
\exp(\pm{\rm i}k_{n}^{\rm h}x),\label{PsiN}
\end{eqnarray}
where the wavenumbers $k_{n}^{\rm e}$ and $k_{n}^{\rm h}$ are given by
\begin{eqnarray}
k_{n}^{\rm e,h}\equiv (2m/\hbar^{2})^{1/2}(E_{\rm F}
-E_{n}+\sigma^{\rm e,h}\varepsilon)^{1/2}, \label{keh}
\end{eqnarray}
and we have defined $\sigma^{\rm e}\equiv 1$, $\sigma^{\rm h}\equiv -1$. The
labels e and h indicate the electron or hole character of the wavefunction.
The index $n$ labels the modes, $\Phi_{n}(y,z)$ is the transverse wavefunction
of the $n$-th mode, and $E_{n}$ its threshold energy:
\begin{eqnarray}
[(p_{y}^{2}+p_{z}^{2})/2m+V(y,z)]\Phi_{n}(y,z)= E_{n}\Phi_{n}(y,z).\label{Phin}
\end{eqnarray}
The eigenfunction $\Phi_{n}$ is normalized to unity, $\int\! {\rm d}y\int\!
{\rm d}z \,|\Phi_{n}|^{2}=1$. With this normalization each wavefunction in the
basis (\ref{PsiN}) carries the same amount of quasiparticle current. The
eigenfunctions in lead ${\rm N}_{1}$ are chosen similarly, but with an
additional phase factor $\exp[-{\rm i}\sigma^{\rm e,h}(eA_{\rm 1}/\hbar)y]$
from the vector potential.

A wave incident on the disordered normal region is described in the basis
(\ref{PsiN}) by a vector of coefficients
\begin{eqnarray}
c_{\rm N}^{\rm in}\equiv\bigl(
c_{\rm e}^{+}({\rm N}_{1}), c_{\rm e}^{-}({\rm N}_{2}),
c_{\rm h}^{-}({\rm N}_{1}), c_{\rm h}^{+}({\rm N}_{2})\bigr).\label{cNin}
\end{eqnarray}
(The mode-index $n$ has been suppressed for simplicity of notation.) The
reflected and transmitted wave has vector of coefficients
\begin{eqnarray}
c_{\rm N}^{\rm out}\equiv\bigl(
c_{\rm e}^{-}({\rm N}_{1}), c_{\rm e}^{+}({\rm N}_{2}),
c_{\rm h}^{+}({\rm N}_{1}), c_{\rm h}^{-}({\rm N}_{2})\bigr).\label{cNout}
\end{eqnarray}
The $s$-matrix $s_{\rm N}$ of the normal region relates these two vectors,
\begin{equation}
c_{\rm N}^{\rm out}=s_{\rm N}^{\vphantom{{\rm in}}}c_{\rm N}^{\rm
in}.\label{sNdef}
\end{equation}
Because the normal region does not couple electrons and holes, this matrix has
the block-diagonal form
\begin{eqnarray}
s_{\rm N}(\varepsilon)=
{\renewcommand{\arraystretch}{0.8}
\left(\begin{array}{cc}
s_{0}(\varepsilon)&0\\
\!0&s_{0}(-\varepsilon)^{\ast}
\end{array}\right)},\,
s_{\rm 0}\equiv{\renewcommand{\arraystretch}{0.6}
\left(\begin{array}{cc}
r_{11}&t_{12}\\t_{21}&r_{22}
\end{array}\right)}.
\label{sN}
\end{eqnarray}
Here $s_{0}$ is the unitary $s$-matrix associated with the single-electron
Hamiltonian ${\cal H}_{0}$. The reflection and transmission matrices
$r(\varepsilon)$ and $t(\varepsilon)$ are $N\times N$ matrices,
$N(\varepsilon)$ being the number of propagating modes at energy $\varepsilon$.
(We assume for simplicity that the number of modes in leads ${\rm N}_{1}$ and
${\rm N}_{2}$ is the same.) The matrix $s_{0}$ is unitary
($s_{0}^{\dagger}s_{0}^{\vphantom{\dagger}}=1$) and satisfies the symmetry
relation $s_{0}(\varepsilon,B)_{ij}=s_{0}(\varepsilon,-B)_{ji}$.

For energies $0<\varepsilon<{\mit\Delta_{0}}$ there are no propagating modes in
the superconductor. We can then define an $s$-matrix for Andreev reflection at
the NS interface which relates the vector of coefficients $\bigl( c_{\rm
e}^{-}({\rm N}_{2}), c_{\rm h}^{+}({\rm N}_{2})\bigr)$ to $\bigl( c_{\rm
e}^{+}({\rm N}_{2}), c_{\rm h}^{-}({\rm N}_{2})\bigr)$. The elements of this
$s$-matrix can be obtained by matching the wavefunctions (\ref{PsiN}) at $x=0$
to the decaying solutions in S of the BdG equation. If terms of order
${\mit\Delta_{0}}/E_{\rm F}$ are neglected (the socalled Andreev approximation
\cite{And64}), the result is simply
\begin{eqnarray}
&&c_{\rm e}^{-}({\rm N}_{2})= \alpha\,{\rm e}^{{\rm i}\phi}c_{\rm h}^{-}({\rm
N}_{2}),\nonumber\\
&&c_{\rm h}^{+}({\rm N}_{2})= \alpha\,{\rm e}^{-{\rm i}\phi}c_{\rm e}^{+}({\rm
N}_{2}),\label{sA}
\end{eqnarray}
where $\alpha\equiv\exp[-{\rm i}\arccos(\varepsilon/{\mit\Delta_{0}})]$.
Andreev reflection transforms an electron mode into a hole mode, without change
of mode index. The transformation is accompanied by a phase shift, which
consists of two parts:
\begin{enumerate}
\item A phase shift $-\arccos(\varepsilon/{\mit\Delta_{0}})$ due to the
penetration of the wavefunction into the superconductor.
\item A phase shift equal to plus or minus the phase of the pair potential in
the superconductor ({\em plus\/} for reflection from hole to electron, {\em
minus\/} for the reverse process).
\end{enumerate}

We can combine the $2N$ linear relations (\ref{sA}) with the $4N$ relations
(\ref{sNdef}) to obtain a set of $2N$ linear relations between the incident
wave in lead ${\rm N}_{1}$ and the reflected wave in the same lead:
\begin{eqnarray}
&&c_{\rm e}^{-}({\rm N}_{1})= s_{\rm ee}^{\vphantom{+}}c_{\rm e}^{+}({\rm
N}_{1})+s_{\rm eh}^{\vphantom{+}}c_{\rm h}^{-}({\rm N}_{1}),\nonumber\\
&&c_{\rm h}^{+}({\rm N}_{1})= s_{\rm he}^{\vphantom{+}}c_{\rm e}^{+}({\rm
N}_{1})+s_{\rm hh}^{\vphantom{+}}c_{\rm h}^{-}({\rm N}_{1}).\label{sdef}
\end{eqnarray}
The four $N\times N$ matrices $s_{\rm ee}$, $s_{\rm hh}$, $s_{\rm eh}$, and
$s_{\rm he}$ form together the scattering matrix $s$ of the whole system for
energies $0<\varepsilon<{\mit\Delta_{0}}$. An electron incident in lead ${\rm
N}_{1}$ is reflected either as an electron (with scattering amplitudes $s_{\rm
ee}$) or as a hole (with scattering amplitudes $s_{\rm he}$). Similarly, the
matrices $s_{\rm hh}$ and $s_{\rm eh}$ contain the scattering amplitudes for
reflection of a hole as a hole or as an electron. After some algebra we find
for these matrices the expressions
\begin{eqnarray}
s_{\rm
ee}^{\vphantom{\ast}}(\varepsilon)&=&r_{11}^{\vphantom{\ast}}(\varepsilon)+
\alpha^{2}t_{12}^{\vphantom{\ast}}(\varepsilon)r_{22}^{\ast}
(-\varepsilon)M_{\rm e}^{\vphantom{\ast}}t_{21}^{\vphantom{\ast}}(\varepsilon),
\label{see}\\
s_{\rm hh}^{\vphantom{\ast}}(\varepsilon)&=&r_{11}^{\ast}(-\varepsilon)+
\alpha^{2}t_{12}^{\ast}(-\varepsilon)r_{22}^{\vphantom{\ast}}
(\varepsilon)M_{\rm
h}^{\vphantom{\ast}}t_{21}^{\ast}(-\varepsilon),\label{shh}\\
s_{\rm eh}^{\vphantom{\ast}}(\varepsilon)&=&\alpha\,{\rm e}^{{\rm
i}\phi}t_{12}^{\vphantom{\ast}}(\varepsilon)M_{\rm
h}^{\vphantom{\ast}}t_{21}^{\ast}(-\varepsilon),\label{seh}\\
s_{\rm he}^{\vphantom{\ast}}(\varepsilon)&=&\alpha\,{\rm e}^{-{\rm
i}\phi}t_{12}^{\ast}(-\varepsilon)M_{\rm
e}^{\vphantom{\ast}}t_{21}^{\vphantom{\ast}}(\varepsilon), \label{she}
\end{eqnarray}
where we have defined the matrices
\begin{eqnarray}
&&M_{\rm e}^{\vphantom{\ast}}\equiv[1-\alpha^{2}
r_{22}^{\vphantom{\ast}}(\varepsilon)r_{22}^{\ast}(-\varepsilon)]^{-1},
\nonumber\\
&&M_{\rm h}^{\vphantom{\ast}}\equiv[1-\alpha^{2}
r_{22}^{\ast}(-\varepsilon)r_{22}(^{\vphantom{\ast}}\varepsilon)]^{-1}.
\label{MeMh}
\end{eqnarray}
One can verify that the $s$-matrix constructed from these four sub-matrices
satisfies unitarity ($s^{\dagger}s=1$) and the symmetry relation
$s(\varepsilon,B,\phi)_{ij}=s(\varepsilon,-B,-\phi)_{ji}$, as required by
quasiparticle-current conservation and by time-reversal invariance,
respectively.

For the linear-response conductance $G_{\rm NS}$ of the NS junction at zero
temperature we only need the $s$-matrix at the Fermi level, i.e.\ at
$\varepsilon=0$. We restrict ourselves to this case and omit the argument
$\varepsilon$ in what follows. We apply the general formula
\cite{Blo82,Lam91,Tak92a}
\begin{equation}
G_{\rm NS}=\frac{2e^{2}}{h}{\rm Tr}\,(1-s_{\rm ee}^{\vphantom{\dagger}}s_{\rm
ee}^{\dagger}+s_{\rm he}^{\vphantom{\dagger}}s_{\rm
he}^{\dagger})=\frac{4e^{2}}{h}{\rm Tr}\,s_{\rm he}^{\vphantom{\dagger}}s_{\rm
he}^{\dagger}.\label{Gdef}
\end{equation}
The second equality follows from unitarity of $s$, which implies $1-s_{\rm
ee}^{\vphantom{\dagger}}s_{\rm ee}^{\dagger}=s_{\rm
eh}^{\vphantom{\dagger}}s_{\rm eh}^{\dagger}=(s_{\rm ee}^{\dagger})^{-1}s_{\rm
he}^{\dagger}s_{\rm he}^{\vphantom{\dagger}}s_{\rm ee}^{\dagger}$, so that
${\rm Tr}\,(1-s_{\rm ee}^{\vphantom{\dagger}}s_{\rm ee}^{\dagger})={\rm
Tr}\,s_{\rm he}^{\vphantom{\dagger}}s_{\rm he}^{\dagger}$. We now substitute
eq.\ (\ref{she}) for $\varepsilon=0$ ($\alpha=-{\rm i}$) into eq.\
(\ref{Gdef}), and obtain the expression
\begin{eqnarray}
G_{\rm NS}=\frac{4e^{2}}{h}{\rm
Tr}\,t_{12}^{\dagger}t_{12}^{\vphantom{\dagger}}
(1+r_{22}^{\ast}r_{22}^{\vphantom{\ast}})^{-1}t_{21}^{\ast}t_{21}^{\rm
T}(1+r_{22}^{\dagger}r_{22}^{\rm T})^{-1},\label{key}
\end{eqnarray}
where $M^{\rm T}\equiv (M^{\ast})^{\dagger}$ denotes the transpose of a matrix.
The advantage of eq.\ (\ref{key}) over eq.\ (\ref{Gdef}) is that the former can
be evaluated by using standard techniques developed for quantum transport in
the normal state, since the only input is the normal-state scattering matrix.
The effects of multiple Andreev reflections are fully incorporated by the two
matrix inversions in eq.\ (\ref{key}).

In the absence of a magnetic field the general formula (\ref{key}) simplifies
considerably. Since the $s$-matrix $s_{0}$ of the normal region is symmetric
for $B=0$, one has $r_{22}^{\vphantom{\rm T}}=r_{22}^{\rm T}$ and
$t_{12}^{\vphantom{\rm T}}=t_{21}^{\rm T}$. Equation (\ref{key}) then takes the
form
\begin{eqnarray}
G_{\rm NS}&=&\frac{4e^{2}}{h}{\rm
Tr}\,t_{12}^{\dagger}t_{12}^{\vphantom{\dagger}}
(1+r_{22}^{\dagger}r_{22}^{\vphantom{\ast}})^{-1}
t_{12}^{\dagger}t_{12}^{\vphantom{\rm T}}(1+r_{22}^{\dagger}
r_{22}^{\vphantom{\rm T}})^{-1}\nonumber\\
&=&\frac{4e^{2}}{h}{\rm
Tr}\left(t_{12}^{\dagger}t_{12}^{\vphantom{\dagger}}
(2-t_{12}^{\dagger}t_{12}^{\vphantom{\dagger}})^{-1}\right)^{2}.
\label{GBzero}
\end{eqnarray}
In the second equality we have used the unitarity relation
$r_{22}^{\dagger}r_{22}^{\vphantom{\ast}}+
t_{12}^{\dagger}t_{12}^{\vphantom{\dagger}}=1$. The trace (\ref{GBzero})
depends only on the eigenvalues of the Hermitian matrix
$t_{12}^{\dagger}t_{12}^{\vphantom{\dagger}}$. We denote these eigenvalues by
$T_{n}$ ($n=1,2,\ldots N$). Since the matrices
$t_{12}^{\dagger}t_{12}^{\vphantom{\dagger}}$,
$t_{12}^{\vphantom{\dagger}}t_{12}^{\dagger}$,
$t_{21}^{\dagger}t_{21}^{\vphantom{\dagger}}$, and
$t_{21}^{\vphantom{\dagger}}t_{21}^{\dagger}$ all have the same set of
eigenvalues, we can omit the indices and write simply $tt^{\dagger}$.  We
obtain the following relation between the conductance and the transmission
eigenvalues:
\begin{equation}
G_{\rm NS}=\frac{4e^{2}}{h}\sum_{n=1}^{N}\frac{T_{n}^{2}}{(2-T_{n})^{2}}.
\label{keyzero}
\end{equation}
This is the central result of ref.\ \cite{Bee92}.

Equation (\ref{keyzero}) holds for an arbitrary transmission matrix $t$, i.e.\
for arbitrary disorder potential. It is the {\em multi-channel\/}
generalization of a formula first obtained by Blonder, Tinkham, and Klapwijk
\cite{Blo82} (and subsequently by Shelankov \cite{She84} and by Za\u{\i}tsev
\cite{Zai84}) for the {\em single-channel\/} case (appropriate for a geometry
such as a planar tunnel barrier, where the different scattering channels are
uncoupled). A formula of similar generality for the normal-metal conductance
$G_{\rm N}$ is the multi-channel Landauer formula
\begin{equation}
G_{\rm N}=\frac{2e^{2}}{h}{\rm Tr}\,tt^{\dagger}\equiv\frac{2e^{2}}{h}
\sum_{n=1}^{N}T_{n}.\label{Landauer}
\end{equation}
In contrast to the Landauer formula, eq.\ (\ref{keyzero}) for the conductance
of an NS junction is a {\em non-linear\/} function of the transmission
eigenvalues $T_{n}$. When dealing with a non-linear multi-channel formula as
eq.\ (\ref{keyzero}), it is of importance to distinguish between the
transmission eigenvalue $T_{n}$ and the modal transmission probability ${\cal
T}_{n}\equiv\sum_{m=1}^{N}|t_{nm}|^{2}$. The former is an eigenvalue of the
matrix $tt^{\dagger}$, the latter a diagonal element of that matrix. The
Landauer formula (\ref{Landauer}) can be written equivalently as a sum over
eigenvalues or as sum over modal transmission probabilities:
\begin{eqnarray}
\frac{h}{2e^{2}}G_{\rm N}=\sum_{n=1}^{N}T_{n}\equiv\sum_{n=1}^{N}{\cal
T}_{n}.\label{Landauer2}
\end{eqnarray}
This equivalence is of importance for (numerical) evaluations of the Landauer
formula, in which one calculates the probability that an electron injected in
mode $n$ is transmitted, and then obtains the conductance by summing over all
modes. The non-linear scattering formula (\ref{keyzero}), in contrast, can not
be written in terms of modal transmission probabilities alone: The off-diagonal
elements of $tt^{\dagger}$ contribute to $G_{\rm NS}$ in an essential way.
Previous attempts to generalize the one-dimensional Blonder-Tinkham-Klapwijk
formula to more dimensions by summing over modal transmission probabilities
(or, equivalently, by angular averaging) were not successful precisely because
only the diagonal elements of $tt^{\dagger}$ were considered.

\section{Three simple applications}

To illustrate the power and generality of the scattering formula
(\ref{keyzero}), we discuss in this section three simple applications to the
ballistic, resonant-tunneling, and diffusive transport regimes \cite{Bee92}.

\subsection{Quantum point contact}

Consider first the case that the normal metal consists of a ballistic
constriction with a normal-state conductance quantized at $G_{\rm
N}=2N_{0}e^{2}/h$ (a {\em quantum point contact\/}). The integer $N_{0}$ is the
number of occupied one-dimensional subbands (per spin direction) in the
constriction, or alternatively the number of transverse modes at the Fermi
level which can propagate through the constriction. Note that $N_{0}\ll N$. An
``ideal'' quantum point contact is characterized by a special set of
transmission eigenvalues, which are equal to either zero or one \cite{Eer91}:
\begin{eqnarray}
T_{n}=\left\{\begin{array}{ll}
1 &\;{\rm if}\; 1\leq n\leq N_{0},\\
0 &\;{\rm if}\; N_{0}<n\leq N,
\end{array}\right.\label{TQPC}
\end{eqnarray}
where the eigenvalues have been ordered from large to small. We emphasize that
eq.\ (\ref{TQPC}) does not imply that the transport through the constriction is
adiabatic. In the case of adiabatic transport, the transmission eigenvalue
$T_{n}$ is equal to the modal transmission probability ${\cal T}_{n}$. In the
absence of adiabaticity there is no direct relation between $T_{n}$ and ${\cal
T}_{n}$. Substitution of eq.\ (\ref{TQPC}) into eq.\ (\ref{keyzero}) yields
\begin{eqnarray}
G_{\rm NS}=\frac{4e^{2}}{h}N_{0}.\label{GQPC}
\end{eqnarray}
The conductance of the NS junction is quantized in units of $4e^{2}/h$. This is
{\em twice\/} the conductance quantum in the normal state, due to the
current-doubling effect of Andreev reflection \cite{Hou91}.

\begin{figure}[tb]
\hspace*{\fill}
\psfig{figure=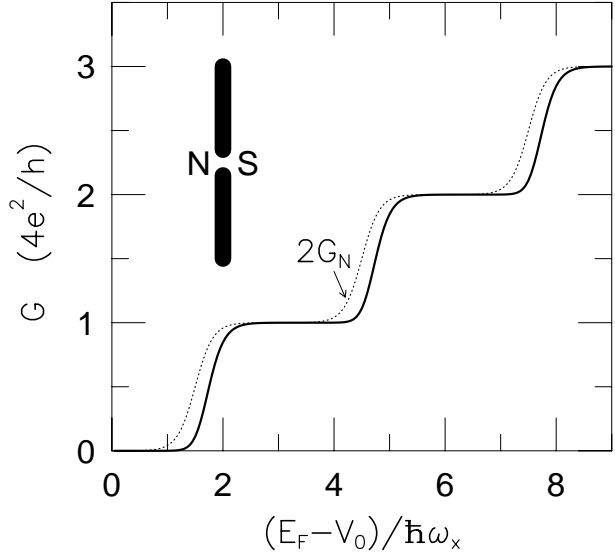,width=
8cm,bbllx=15pt,bblly=125pt,bburx=560pt,bbury=626pt}
\hspace*{\fill}
\medskip\caption[]{Solid curve: Conductance $G_{\rm NS}$ versus Fermi energy of a
quantum point contact between a normal and a superconducting reservoir (shown
schematically in the inset). The dotted curve is twice the conductance $G_{\rm
N}$ for the case of two normal reservoirs \protect\cite{But90}. The
constriction is defined by the 2D saddle-point potential
$V(x,y)=V_{0}-{\textstyle\frac{1}{2}}m\omega_{x}^{2}x^{2}+
{\textstyle\frac{1}{2}}m\omega_{y}^{2}y^{2}$, with $\omega_{y}/\omega_{x}=3$;
$G_{\rm NS}$ is calculated from eq.\ (\protect\ref{keyzero}), with
$T_{n}=[1+\exp(-2\pi\varepsilon_{n}/\hbar\omega_{x})]^{-1}$,
$\varepsilon_{n}\equiv E_{\rm
F}-V_{0}-(n-{\textstyle\frac{1}{2}})\hbar\omega_{y}$. {\em (From ref.\
\protect\cite{Bee92}.)}
\label{sqpcn}}
\end{figure}

In the classical limit $N_{0}\rightarrow\infty$ we recover the well-known
result $G_{\rm NS}=2G_{\rm N}$ for a {\em classical\/} ballistic point contact
\cite{Blo82,She84,Zai80}. In the quantum regime, however, the simple
factor-of-two enhancement only holds for the conductance plateaus, where eq.\
(\ref{TQPC}) applies, and not to the transition region between two subsequent
plateaus of quantized conductance. To illustrate this, we compare in fig.\
\ref{sqpcn} the conductances $G_{\rm NS}$ and $2G_{\rm N}$ for B\"{u}ttiker's
model \cite{But90} of a saddle-point constriction in a two-dimensional electron
gas. Appreciable differences appear in the transition region, where $G_{\rm
NS}$ lies below twice $G_{\rm N}$. This is actually a rigorous inequality,
which follows from eqs.\ (\ref{keyzero}) and (\ref{Landauer}) for {\em
arbitrary\/} transmission matrix:
\begin{eqnarray}
G_{\rm NS}\leq 2G_{\rm N},\;\forall\, t.\label{Gleg2G0}
\end{eqnarray}

\subsection{Quantum dot}

Consider next a small confined region (of dimensions comparable to the Fermi
wavelength), which is weakly coupled by tunnel barriers to two electron
reservoirs. We assume that transport through this {\em quantum dot\/} occurs
via resonant tunneling through a single bound state. Let $\varepsilon_{\rm
res}$ be the energy of the resonant level, relative to the Fermi level in the
reservoirs, and let $\gamma_{1}/\hbar$ and $\gamma_{2}/\hbar$ be the tunnel
rates through the two barriers. We denote $\gamma\equiv\gamma_{1}+\gamma_{2}$.
If $\gamma\ll\Delta E$ (with $\Delta E$ the level spacing in the quantum dot),
the conductance $G_{\rm N}$ in the case of non-interacting electrons has the
form
\begin{eqnarray}
\frac{h}{2e^{2}}G_{\rm N}=\frac{\gamma_{1}\gamma_{2}}{\varepsilon_{\rm
res}^{2}+{{\textstyle\frac{1}{4}}}\gamma^{2}}\equiv T_{\rm BW},\label{G0BW}
\end{eqnarray}
with $T_{\rm BW}$ the Breit-Wigner transmission probability at the Fermi level.
The normal-state transmission matrix $t_{12}(\varepsilon)$ which yields this
conductance has matrix elements \cite{But88}
\begin{eqnarray}
t_{12}(\varepsilon)=U_{1}\tau(\varepsilon)U_{2},
\;\tau(\varepsilon)_{nm}\equiv\frac{\sqrt{\gamma_{1n}\gamma_{2m}}}
{\varepsilon-\varepsilon_{\rm res}+{\textstyle\frac{1}{2}}{\rm
i}\gamma},\label{sBW}
\end{eqnarray}
where $\sum_{n}\gamma_{1n}\equiv\gamma_{1}$,
$\sum_{n}\gamma_{2n}\equiv\gamma_{2}$, and $U_{1}$, $U_{2}$ are two unitary
matrices (which need not be further specified).

\begin{figure}[tb]
\hspace*{\fill}
\psfig{figure=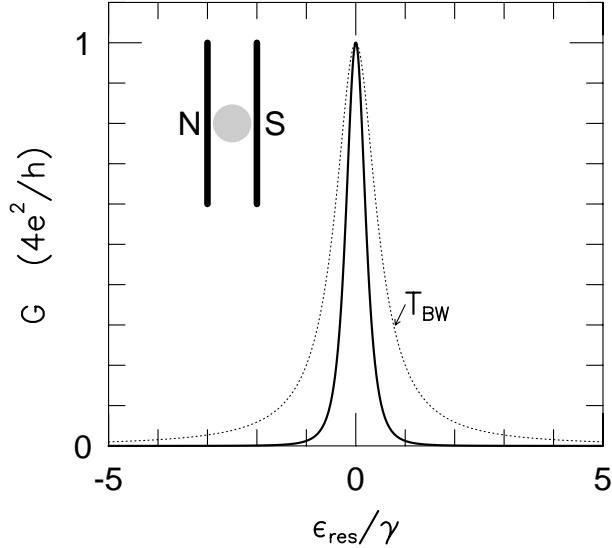,width=
8cm,bbllx=17pt,bblly=125pt,bburx=568pt,bbury=626pt}
\hspace*{\fill}
\medskip\caption[]{Conductance versus energy of the resonant level, from eq.\
(\ref{GBW}) for the case of equal tunnel barriers (solid curve). The dotted
curve is the Breit-Wigner transmission probability (\protect\ref{G0BW}). The
inset shows schematically the normal-metal --- quantum-dot --- superconductor
junction.
\label{sqdotn}}
\end{figure}

Let us now investigate how the conductance (\ref{G0BW}) is modified if one of
the two reservoirs is in the superconducting state. The transmission matrix
product $t_{12}^{\vphantom{\dagger}}t_{12}^{\dagger}$ (evaluated at the Fermi
level $\varepsilon=0$) following from eq.\ (\ref{sBW}) is
\begin{eqnarray}
t_{12}^{\vphantom{\dagger}}t_{12}^{\dagger}=
U_{1}^{\vphantom{\dagger}}MU_{1}^{\dagger},
\;M_{nm}^{\vphantom{\dagger}}\equiv\frac{T_{\rm
BW}}{\gamma_{1}}\sqrt{\gamma_{1n}\gamma_{1m}}.\label{ttBW}
\end{eqnarray}
Its eigenvalues are
\begin{eqnarray}
T_{n}=\left\{\begin{array}{ll}
T_{\rm BW} &\;{\rm if}\; n=1,\\
0 &\;{\rm if}\; 2\leq n\leq N.
\end{array}\right.\label{TBW}
\end{eqnarray}
Substitution into eq.\ (\ref{keyzero}) yields the conductance
\begin{equation}
G_{\rm NS}=\frac{4e^{2}}{h}\left( \frac{T_{\rm BW}}{2-T_{\rm BW}}\right) ^{2}
=\frac{4e^{2}}{h}\left(\frac{2\gamma_{1}\gamma_{2}}{4\varepsilon_{\rm
res}^{2}+\gamma_{1}^{2}+\gamma_{2}^{2}}\right) ^{2}.\label{GBW}
\end{equation}
The conductance on resonance ($\varepsilon_{\rm res}=0$) is maximal in the case
of equal tunnel rates ($\gamma_{1}=\gamma_{2}$), and is then equal to
$4e^{2}/h$ --- independent of $\gamma$. The lineshape for this case is shown in
fig.\ \ref{sqdotn} (solid curve). It differs substantially from the Lorentzian
lineshape (\ref{G0BW}) of the Breit-Wigner formula (dotted curve).

The amplitude and lineshape of the conductance resonance (\ref{GBW}) does not
depend on the relative magnitude of the resonance width $\gamma$ and the
superconducting energy gap ${\mit\Delta_{0}}$. This is in contrast to the
supercurrent resonance in a superconductor --- quantum dot --- superconductor
Josephson junction, which depends sensitively on the ratio
$\gamma/{\mit\Delta_{0}}$ \cite{Gla89,SQUID91}. The difference can be traced to
the fact that the conductance (in the zero-temperature, zero-voltage limit) is
strictly a Fermi-level property, whereas all states within ${\mit\Delta_{0}}$
of the Fermi level contribute to the Josephson effect. (For an extension of
eq.\ (\ref{GBW}) to finite voltages, see ref.\ \cite{Khl93}.) Since we have
assumed non-interacting quasiparticles, the above results apply to a quantum
dot with a small charging energy $U$ for double occupancy of the resonant
state. Devyatov and Kupriyanov \cite{Dev90}, and Hekking et al.\ \cite{Hek93a},
have studied the influence of Coulomb repulsion on resonant tunneling through
an NS junction, in the temperature regime $k_{\rm B}T\gg\gamma$ where the
resonance is thermally broadened. The extension to the low-temperature regime
of an intrinsically broadened resonance remains to be investigated.

\subsection{Disordered junction}

We now turn to the regime of diffusive transport through a disordered point
contact or microbridge between a normal and a superconducting reservoir. The
model considered is that of an NS junction containing a disordered normal
region of length $L$ much greater than the mean free path $l$ for elastic
impurity scattering, but much smaller than the localization length $Nl$. We
calculate the average conductance of the junction, averaged over an ensemble of
impurity configurations. We begin by parameterizing the transmission eigenvalue
$T_{n}$ in terms of a channel-dependent localization length $\zeta_{n}$:
\begin{equation}
T_{n}=\frac{1}{\cosh^{2}(L/\zeta_{n})}.\label{xip}
\end{equation}
A fundamental result in quantum transport is that the inverse localization
length is {\em uniformly\/} distributed between $0$ and $1/\zeta_{\rm
min}\simeq 1/l$ for $l\ll L\ll Nl$ \cite{Dor84,Imr86,Pen92,Naz94}. One can
therefore write
\begin{equation}
\frac{\left\langle\sum_{n=1}^{N}f(T_{n})\right\rangle}
{\left\langle\sum_{n=1}^{N}T_{n}\right\rangle}=\frac{\int_{0}^{L/\zeta_{\rm
min}}\! {\rm d}x \, f(\cosh^{-2}x)}{\int_{0}^{L/\zeta_{\rm min}}\! {\rm d}x \,
\cosh^{-2}x}=\int_{0}^{\infty}\! {\rm d}x\, f(\cosh^{-2}x),\label{avf}
\end{equation}
where $\langle\ldots\rangle$ indicates the ensemble average and $f(T)$ is an
arbitrary function of the transmission eigenvalue such that $f(T)\rightarrow 0$
for $T\rightarrow 0$. In the second equality in eq.\ (\ref{avf}) we have used
that $L/\zeta_{\rm min}\simeq L/l\gg 1$ to replace the upper integration limit
by $\infty$.

Combining eqs.\ (\ref{keyzero}), (\ref{Landauer}), and (\ref{avf}), we find
\begin{equation}
\langle G_{\rm NS}\rangle=2\langle G_{\rm N}\rangle\int_{0}^{\infty}\! {\rm
d}x\,\left( \frac{\cosh^{-2}x}{2-\cosh^{-2}x}\right) ^{2}=\langle G_{\rm
N}\rangle .\label{Gav}
\end{equation}
We conclude that --- although $G_{\rm NS}$ according to eq.\ (\ref{keyzero}) is
of {\em second\/} order in the transmission eigenvalues $T_{n}$ --- the
ensemble average $\langle G_{\rm NS}\rangle$ is of {\em first\/} order in
$l/L$. The resolution of this paradox is that the $T$'s are not distributed
uniformly, but are either exponentially small (closed channels) or of order
unity (open channels) \cite{Imr86}. Hence the average of $T_{n}^{2}$ is of the
same order as the average of $T_{n}$. Off-diagonal elements of the transmission
matrix $tt^{\dagger}$ are crucial to arrive at the result (\ref{Gav}). Indeed,
if one would evaluate eq.\ (\ref{keyzero}) with the transmission eigenvalues
$T_{n}$ replaced by the modal transmission probabilities ${\cal T}_{n}$, one
would find a totally wrong result: Since ${\cal T}_{n}\simeq l/L\ll 1$, one
would find $G_{\rm NS}\simeq (l/L)G_{\rm N}$ --- which underestimates the
conductance of the NS junction by the factor $L/l$.

Previous work \cite{And66,Art79} had obtained the equality of $G_{\rm NS}$ and
$G_{\rm N}$ from {\em semiclassical\/} equations of motion, as was appropriate
for macroscopic systems which are large compared to the normal-metal
phase-coherence length $l_{\phi}$. The present derivation, in contrast, is
fully quantum mechanical. It applies to the ``mesoscopic'' regime $L<l_{\phi}$,
in which transport is phase coherent. Takane and Ebisawa \cite{Tak92b} have
studied the conductance of a disordered phase-coherent NS junction by numerical
simulation of a two-dimensional tight-binding model. They found $\langle G_{\rm
NS}\rangle =\langle G_{\rm N}\rangle$ within numerical accuracy for $l\ll L\ll
Nl$, in agreement with eq.\ (\ref{Gav}).

If the condition $L\ll Nl$ is relaxed, differences between $\langle G_{\rm
NS}\rangle$ and $\langle G_{\rm N}\rangle$ appear. To lowest order in $L/Nl$,
the difference is a manifestation of the {\em weak-localization\/} effect, as
we discuss in the following section.

\section{Weak localization}

An NS junction shows an {\em enhanced\/} weak-localization effect, in
comparison with the normal state \cite{Bee92}. The origin of the enhancement
can be understood in a simple way, as follows.

We return to the parameterization $T_{n}\equiv 1/\cosh^{2}(L/\zeta_{n})$
introduced in eq.\ (\ref{xip}), and define the density of localization lengths
$\rho(\zeta,L)\equiv\langle\sum_{n}
\delta(\zeta-\zeta_{n})\rangle_{L}$. The subscript $L$ refers to the length of
the disordered region. Using the identity $\cosh 2x=2\cosh^{2}x-1$, the
ensemble-average of eq.\ (\ref{keyzero}) becomes
\begin{equation}
\langle G_{\rm NS}\rangle_{L} =\frac{4e^{2}}{h}\int_{0}^{\infty}\!
{\rm d}\zeta\,\rho(\zeta,L)\cosh^{-2}(2L/\zeta).\label{GNSzeta}
\end{equation}
In the same parameterization, one has
\begin{equation}
\langle G_{\rm N}\rangle_{L} =\frac{2e^{2}}{h}\int_{0}^{\infty}\!
{\rm d}\zeta\,\rho(\zeta,L)\cosh^{-2}(L/\zeta).\label{GNzeta}
\end{equation}
In the ``open-channel approximation'' \cite{Sto91}, the integrals over $\zeta$
are restricted to the range $\zeta >L$ of localization lengths greater than the
length of the conductor. In this range the density $\rho(\zeta,L)$ is
approximately independent of $L$. The whole $L$-dependence of the integrands in
eqs.\ (\ref{GNSzeta}) and (\ref{GNzeta}) lies then in the argument of the
hyperbolic cosine, so that
\begin{equation}
\langle G_{\rm NS}\rangle_{L}=2\langle G_{\rm
N}\rangle_{2L}.\label{openchannel}
\end{equation}
This derivation formalizes the intuitive notion that Andreev reflection at an
NS interface effectively doubles the length of the normal-metal conductor
\cite{Tak92b}.

Consider now the geometry $W\ll L$ relevant for a microbridge. In the normal
state one has
\begin{equation}
\langle G_{\rm N}\rangle=(W/L)\sigma_{\rm Drude}-\delta G_{\rm
N},\label{weaklocaN}
\end{equation}
where $\sigma_{\rm Drude}$ is the classical Drude conductivity. The
$L$-independent term $\delta G_{\rm N}$ is the weak-localization correction,
given by \cite{Mel91} $\delta G_{N}=\frac{2}{3}\,e^{2}/h$. Equation
(\ref{openchannel}) then implies that
\begin{equation}
\langle G_{\rm NS}\rangle
=(W/L)\sigma_{\rm Drude}-\delta G_{\rm NS},\label{weaklocaNS}
\end{equation}
with $\delta G_{\rm NS}=2\,\delta G_{\rm N}$. We conclude that Andreev
reflection increases the weak-localization correction, by a factor of two
according to this qualitative argument \cite{Bee92}. A rigorous theory
\cite{Bee94,Mac94,Tak94} of weak localization in an NS microbridge shows that
the increase is actually somewhat less than a factor of two,\footnote{
Equation (\protect\ref{weaklocaNSN}) follows from the general formula $\delta
A=\frac{1}{4}a(1)+\int_{0}^{\infty}\!{\rm d}x\,(4x^{2}+
\pi^{2})^{-1}a(\cosh^{-2}x)$ for the weak-localization correction in a wire
geometry, where $A$ is an arbitrary transport property of the form
$A=\sum_{n}a(T_{n})$.}
\begin{equation}
\delta G_{\rm NS}=(2-8\pi^{-2})\,e^{2}/h=1.78\,\delta G_{\rm
N}.\label{weaklocaNSN}
\end{equation}

As pointed out in ref.\ \cite{Mar93}, the enhancement of weak localization in
an NS junction can be observed experimentally as a {\em dip\/} in the
differential conductance $G_{\rm NS}(V)=\partial I/\partial V$ around zero
voltage. The dip occurs because an applied voltage destroys the enhancement of
weak localization by Andreev reflection, thereby increasing the conductance by
an amount
\begin{equation}
\delta G_{\rm NS}-\delta G_{\rm N}\approx 0.5\,e^{2}/h\label{dipsize}
\end{equation}
at zero temperature. [At finite temperatures, we expect a reduction of the size
of dip by a factor\footnote{
The reduction factor $(L_{\rm c}/L)^{2}$ for the size of the conductance dip
when $W<L_{\rm c}<L$ is estimated as follows: Consider the wire as consisting
of $L/L_{\rm c}$ phase-coherent segments of length $L_{\rm c}$ in series. The
first segment, adjacent to the superconductor, has a conductance dip $\delta
G_{1}\simeq e^{2}/h$, while the other segments have no conductance dip. The
resistance $R_{1}$ of a single segment is a fraction $L_{\rm c}/L$ of the total
resistance $R$ of the wire. Since $\delta G/G=-\delta R/R=-\delta R_{1}/R$ and
$\delta R_{1}=-R_{1}^{2}\delta G_{1}\simeq -(L_{\rm c}/L)^{2}R^{2}e^{2}/h$, we
find $\delta G\simeq(L_{\rm c}/L)^{2}e^{2}/h$.
}
$(L_{\rm c}/L)^{2}$, where $L_{\rm c}={\rm min}\,(l_{\phi},\sqrt{\hbar D/k_{\rm
B}T})$ is the length over which electrons and holes remain phase coherent.] We
emphasize that in the normal state, weak localization can {\em not\/} be
detected in the current--voltage characteristic. The reason why a dip occurs in
$G_{\rm NS}(V)$ and not in $G_{\rm N}(V)$ is that an applied voltage (in
contrast to a magnetic field) does not break time-reversal symmetry --- but
only affects the phase coherence between the electrons and the
Andreev-reflected holes (which differ in energy by up to $2eV$). The width
$V_{\rm c}$ of the conductance dip is of the order of the Thouless energy
$E_{\rm c}\equiv\pi\hbar D/L^{2}$ (with $D$ the diffusion coefficient of the
junction; $L$ should be replaced by $L_{\rm c}$ if $L>L_{\rm c}$). This energy
scale is such that an electron and a hole acquire a phase difference of order
$\pi$ on traversing the junction. The energy $E_{\rm c}$ is much smaller than
the superconducting energy gap ${\mit\Delta_{0}}$, provided $L\gg\xi$ (with
$\xi\simeq(\hbar D/{\mit\Delta_{0}})^{1/2}$ the superconducting coherence
length in the dirty-metal limit). The separation of energy scales is important,
in order to be able to distinguish experimentally the current due to Andreev
reflection below the energy gap from the quasi-particle current above the
energy gap.

\begin{figure}[tb]
\hspace*{\fill}
\psfig{figure=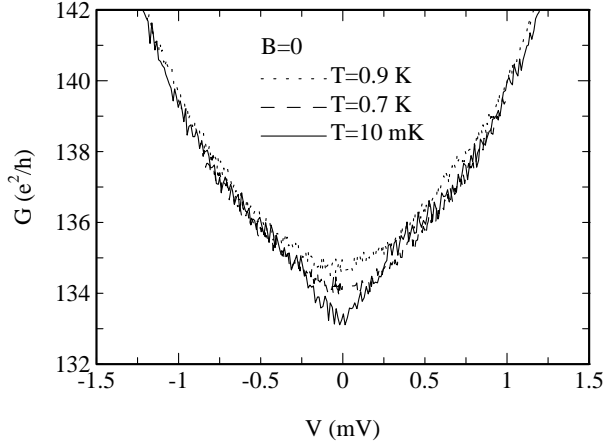,width=8cm}
\hspace*{\fill}
\medskip\caption[]{Differential conductance as a function of applied voltage at three
different temperatures. Experimental data by Lenssen et al.\ for a
two-dimensional electron gas with superconducting contacts. The dip around zero
voltage, which is superimposed on the broad minimum at the lowest temperature,
is attributed to the enhancement of weak localization by Andreev reflection.
{\em (From ref.\ \protect\cite{Len94}.)}
\label{Lenssen}}
\end{figure}

The first measurement of the conductance dip predicted in ref.\ \cite{Mar93}
has been reported recently by Lenssen et al.\ \cite{Len94}. The system studied
consists of the two-dimensional electron gas in a GaAs/AlGaAs heterostructure
with Sn/Ti superconducting contacts ($W=10\,\mu{\rm m}$, $L=0.8\,\mu{\rm m}$).
No supercurrent is observed, presumably because $l_{\phi}\simeq 0.4\,\mu{\rm
m}$ is smaller than $L$. (The phase-coherence length $l_{\phi}$ is estimated
from a conventional weak-localization measurement in a magnetic field.) The
data for the differential conductance is reproduced in fig.\ \ref{Lenssen}. At
the lowest temperatures (10 mK) a rather small and narrow conductance dip
develops, superimposed on a large and broad conductance minimum. The size of
the conductance dip is about $2\,e^{2}/h$. Since in the experimental geometry
$W>L>l_{\phi}$, and there are two NS interfaces, we would expect a dip of order
$2(W/l_{\phi})(l_{\phi}/L)^2\times 0.5\,e^{2}/h\simeq 6\,e^{2}/h$, simply by
counting the number of phase-coherent segments adjacent to the superconductor.
This is three times as large as observed, but the presence of a tunnel barrier
at the NS interface might easily account for this discrepancy. (The Schottky
barrier at the interface between a semiconductor and superconductor presents a
natural origin for such a barrier.) The conductance dip has width $V_{\rm
c}\simeq 0.25\,{\rm mV}$, which is less than the energy gap
${\mit\Delta_{0}}=0.56\,{\rm meV}$ of bulk Sn --- but not by much. Experiments
with a larger separation of energy scales are required for a completely
unambiguous identification of the phenomenon.

\begin{figure}[tb]
\hspace*{\fill}
\psfig{figure=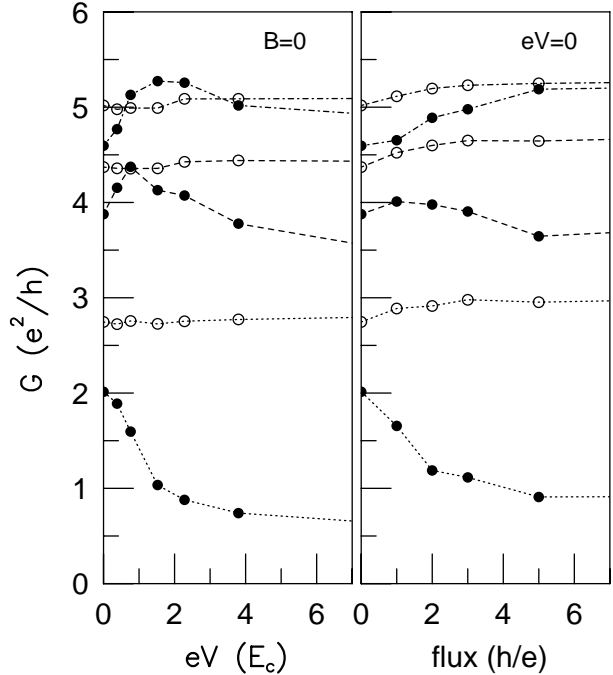,width=8cm}
\hspace*{\fill}
\medskip\caption[]{Voltage and magnetic field dependence of $G_{\rm NS}$ (filled
circles) and $G_{\rm N}$ (open circles). Numerical simulation of a disordered
normal region ($L/W=4.8$, $l/L=0.12$, $N=14$) in series with a tunnel barrier
(transmission probability per mode $\Gamma$; dotted lines: $\Gamma=0.2$;
dashed: $\Gamma=0.6$; dash-dotted: $\Gamma=1$). Note the crossover from a {\em
dip\/} (weak localization) to a {\em peak\/} (reflectionless tunneling) in
$G_{\rm NS}$ on reducing $\Gamma$. {\em (From ref.\ \protect\cite{Mar93}.)}
\label{marmo_fig2}}
\end{figure}

An essential requirement for the appearance of a dip in the differential
conductance is a high probability for Andreev reflection at the NS boundary.
This is illustrated in fig.\ \ref{marmo_fig2}, which shows the results of
numerical simulations \cite{Mar93} of transport through a disordered normal
region connected via a tunnel barrier to a superconductor. The tunnel barrier
is characterized by a transmission probability per mode $\Gamma$. The
dash-dotted lines refer to an ideal interface ($\Gamma=1$), and show the
conductance {\em dip\/} due to weak localization, discussed above. For
$\Gamma\simeq 0.2$--$0.4$ the data for $G_{\rm NS}$ (filled circles) shows a
crossover\footnote{
The crossover is accompanied by an ``overshoot'' around $eV\approx E_{\rm c}$,
indicating the absence of an ``excess current'' (i.e.\ the linear $I$--$V$
characteristic for $eV\gg E_{\rm c}$ extrapolates back through the origin). We
do not have an analytical explanation for the overshoot.
}
to a conductance {\em peak\/}. This is the phenomenon of {\em reflectionless
tunneling\/}, discussed in the following section.

\section{Reflectionless tunneling}

\begin{figure}[tb]
\hspace*{\fill}
\psfig{figure=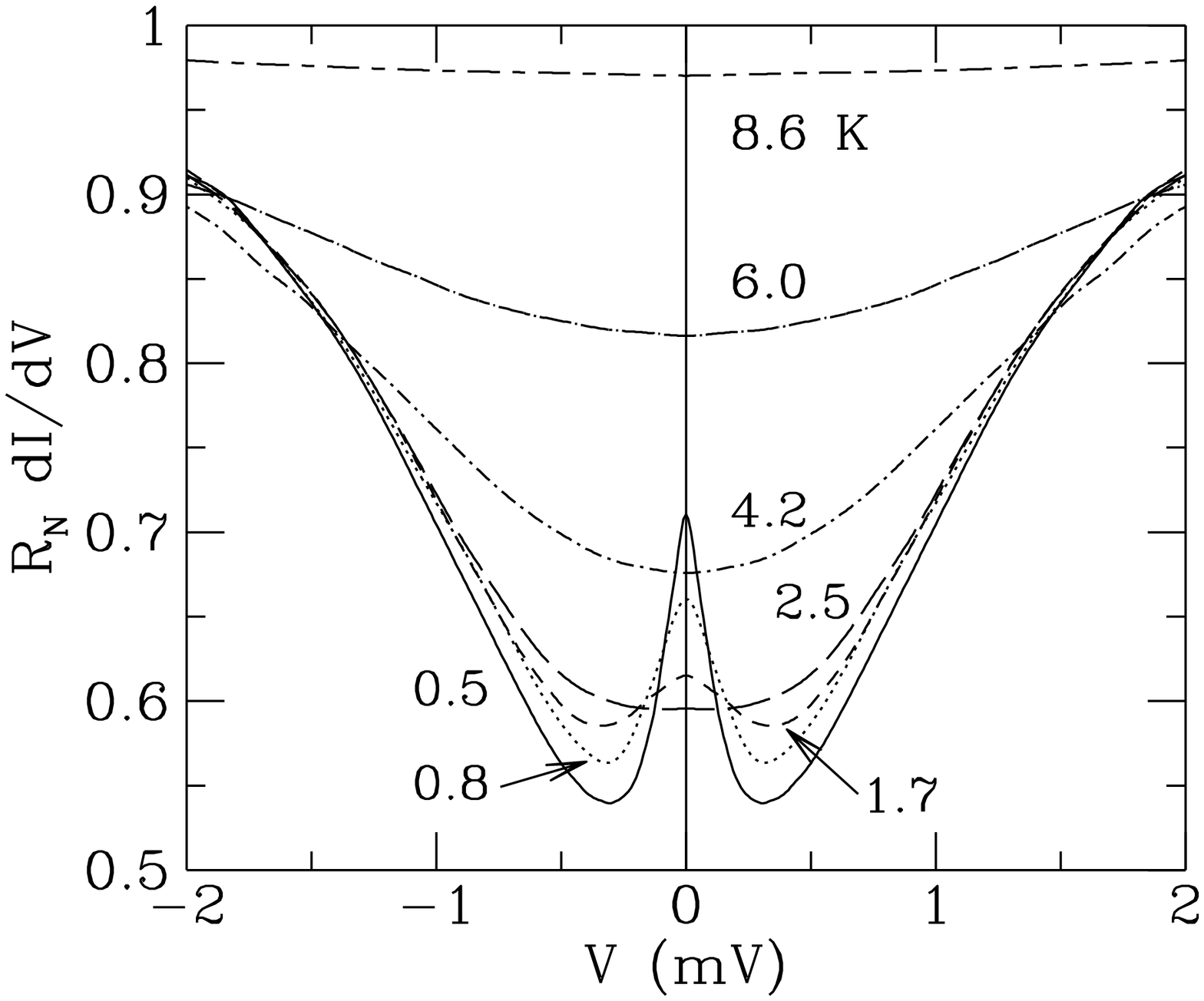,width=
8cm,bbllx=19pt,bblly=190pt,bburx=552pt,bbury=612pt}
\hspace*{\fill}
\medskip\caption[]{Differential conductance (normalized by the normal-state resistance
$R_{\rm N}=0.27\,\Omega$) as a function of applied voltage at seven different
temperatures. Experimental data by Kastalsky et al.\ for a Nb--InGaAs junction.
Note the difference with fig.\ \protect\ref{Lenssen}: A peak rather than a dip
develops at the lowest temperatures, and the size of the peak
($0.6\,\Omega^{-1}\simeq 1.5\cdot 10^{4}\,e^{2}/h$) is four orders of magnitude
greater. The width of the peak is comparable to the width of the dip in fig.\
\protect\ref{Lenssen}. {\em (From ref.\ \protect\cite{Kas91}.)}
\label{Kastalsky}}
\end{figure}

In 1991, Kastalsky et al. \cite{Kas91} discovered a large and narrow peak in
the differential conductance of a Nb--InGaAs junction. We reproduce their data
in fig.\ \ref{Kastalsky}. (A similar peak is observed as a function of magnetic
field.) Since then a great deal of experimental
\cite{Ngu92,Man92,Agr92,Xio93,Len94b,Bak94,Mag94}, numerical
\cite{Mar93,Tak93a}, and analytical work \cite{Wee92,Tak92c,Vol93,Hek93,Bee94b}
has been done on this effect. Here we focus on the explanation in terms of {\em
disorder-induced opening of tunneling channels\/} \cite{Naz94,Bee94b}, which is
the most natural from the view point of the scattering formula (\ref{keyzero}),
and which we feel captures the essence of the effect. Equivalently, the
conductance peak can be explained in terms of a non-equilibrium proximity
effect, which is the preferred explanation in a Green's function formulation of
the problem \cite{Vol93,Zai90,Vol92a,Vol92b}. We begin by reviewing the
numerical work \cite{Mar93}.

\subsection{Numerical simulations}

A sharp peak in the conductance around $V,B=0$ is evident in the numerical
simulations for $\Gamma=0.2$ (dotted lines in fig.\ \ref{marmo_fig2}). While
$G_{\rm N}$ depends only weakly on $B$ and $V$ in this range (open circles),
$G_{\rm NS}$ drops abruptly (filled circles). The width of the conductance peak
in $B$ and $eV$ is respectively of order $B_{\rm c}=h/eLW$ (one flux quantum
through the normal region) and $eV_{\rm c}=\pi\hbar D/L^{2}\equiv E_{\rm c}$
(the Thouless energy). The width of the peak is the same as the width of the
conductance dip due to weak localization, which occurs for larger barrier
transparencies. The size of the peak is much greater than the dip, however.

It is instructive to first discuss the {\em classical\/} resistance $R_{\rm
NS}^{\rm class}$ of the NS junction. The basic approximation in $R_{\rm
NS}^{\rm class}$ is that currents rather than amplitudes are matched at the NS
interface \cite{And66}. The result is
\begin{equation}
R_{\rm NS}^{\rm class}=(h/2Ne^{2})\left[L/l+2\Gamma^{-2}+{\cal
O}(1)\right].\label{RNSclass}
\end{equation}
The contribution from the barrier is $\propto\Gamma^{-2}$ because tunneling
into a superconductor is a two-particle process \cite{She80}: Both the incident
electron and the Andreev-reflected hole have to tunnel through the barrier (the
net result being the addition of a Cooper pair to the superconducting
condensate \cite{And64}). Equation (\ref{RNSclass}) is to be contrasted with
the classical resistance $R_{\rm N}^{\rm class}$ in the normal state,
\begin{equation}
R_{\rm N}^{\rm class}=(h/2Ne^{2})\left[L/l+\Gamma^{-1}+{\cal
O}(1)\right],\label{GNclass}
\end{equation}
where the contribution of a resistive barrier is $\propto\Gamma^{-1}$. In the
absence of a tunnel barrier (i.e.\ for $\Gamma=1$), $R_{\rm NS}^{\rm
class}=R_{\rm N}^{\rm class}$ for $L\gg l$, in agreement with refs.\
\cite{And66,Art79}. Let us now see how these classical results compare with the
simulations \cite{Mar93}.

\begin{figure}[tb]
\hspace*{\fill}
\psfig{figure=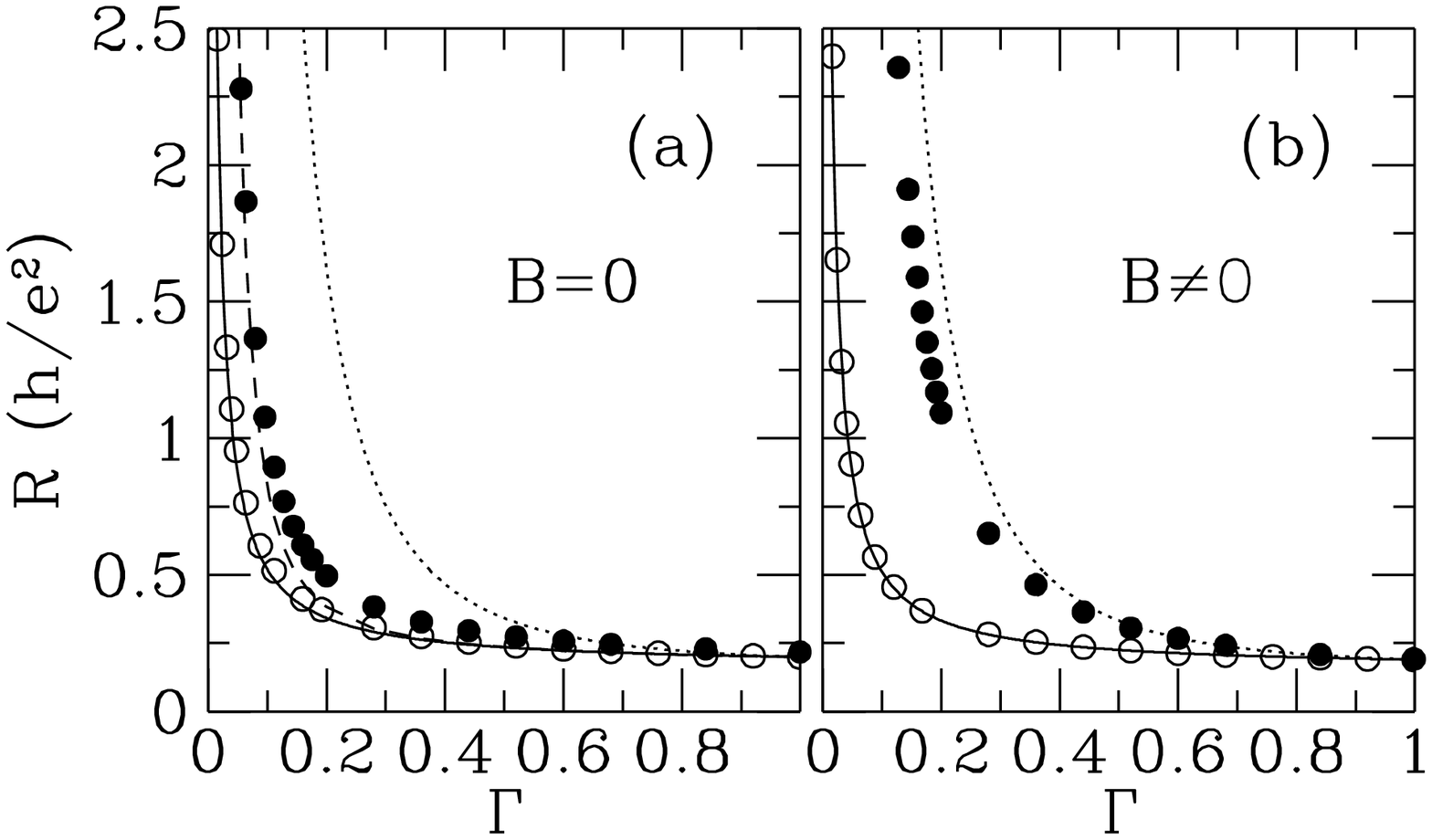,width=
10cm,bbllx=38pt,bblly=183pt,bburx=584pt,bbury=505pt}
\hspace*{\fill}
\medskip\caption[]{Filled circles: Numerically calculated resistance $R_{\rm NS}$ of a
disordered NS junction, versus the transmission probability per mode $\Gamma$
of the tunnel barrier at the NS interface; Open circles: Resistance $R_{\rm N}$
of the same junction in the normal state; (a) is for zero magnetic field, (b)
is for a flux of $10\,h/e$ through the disordered region. The dotted and solid
curves are the classical eqs.\ (\protect\ref{RNSclass}) and
(\protect\ref{GNclass}). The dashed curve is the theory of ref.\
\protect\cite{Vol93}, which for $\Gamma\gg l/L\approx 0.12$ coincides with eq.\
(\protect\ref{RNSzeroB}). {\em (From ref.\ \protect\cite{Mar93}.)}
\label{marmo_fig1}}
\end{figure}

In fig.\ \ref{marmo_fig1} we show the resistance (at $V=0$) as a function of
$\Gamma$ in the absence and presence of a magnetic field. (The parameters of
the disordered region are the same as for fig.\ \ref{marmo_fig2}.) There is
good agreement with the classical eqs.\ (\ref{RNSclass}) and (\ref{GNclass})
for a magnetic field corresponding to 10 flux quanta through the disordered
segment (fig.\ \ref{marmo_fig1}b). For $B=0$, however, the situation is
different (fig.\ \ref{marmo_fig1}a). The normal-state resistance (open circles)
still follows approximately the classical formula (solid curve). (Deviations
due to weak localization are noticeable, but small on the scale of the figure.)
In contrast, the resistance of the NS junction (filled circles) lies much below
the classical prediction (dotted curve). The numerical data shows that for
$\Gamma\gg l/L$ one has approximately
\begin{equation}
R_{\rm NS}(B=0,V=0)\approx R_{\rm N}^{\rm class},\label{RNSzeroB}
\end{equation}
which for $\Gamma\ll 1$ is much smaller than $R_{\rm NS}^{\rm class}$. This is
the phenomenon of {\em reflectionless tunneling}: In fig.\ \ref{marmo_fig1}a
the barrier contributes to $R_{\rm NS}$ in order $\Gamma^{-1}$, just as for
single-particle tunneling, and not in order $\Gamma^{-2}$, as expected for
two-particle tunneling. It is as if the Andreev-reflected hole is not reflected
by the barrier. The interfering trajectories responsible for this effect were
first identified by Van Wees et
al.\ \cite{Wee92}. The numerical data of fig.\ \ref{marmo_fig1}a is in good
agreement with the Green's function calculation of Volkov, Za\u{\i}tsev, and
Klapwijk \cite{Vol93} (dashed curve). Both these papers have played a crucial
role in the understanding of the effect. The scaling theory reviewed below
\cite{Bee94b} is essentially equivalent to the Green's function calculation,
but has the advantage of explicitly demonstrating how the opening of tunneling
channels on increasing the length $L$ of the disordered region induces a
transition from a $\Gamma^{-2}$ dependence to a $\Gamma^{-1}$ dependence when
$L\simeq l/\Gamma$.

\subsection{Scaling theory}

We use the parameterization
\begin{equation}
T_{n}=\frac{1}{\cosh^{2}x_{n}},\label{Txdef}
\end{equation}
similar to eq.\ (\ref{xip}), but now with a dimensionless variable
$x_{n}\in[0,\infty)$. The density of the $x$-variables, for a length $L$ of
disordered region, is denoted by
\begin{equation}
\rho(x,L)=\langle{\textstyle\sum_{n}}\delta(x-x_{n})\rangle_{L}.
\label{rhoxdef}
\end{equation}
For $L=0$, i.e.\ in the absence of disorder, we have the initial condition
imposed by the barrier,
\begin{equation}
\rho(x,0)=N\delta(x-x_{0}),\label{rhox0}
\end{equation}
with $\Gamma=1/\cosh^{2}x_{0}$. The scaling theory describes how $\rho(x,L)$
evolves with increasing $L$. This evolution is governed by the equation
\begin{equation}
\frac{\partial}{\partial s}\rho(x,s)=-\frac{1}{2N} \frac{\partial}{\partial
x}\rho(x,s)\frac{\partial}{\partial x}
\int_{0}^{\infty}\!\!{\rm d}x'\,
\rho(x',s)\ln|\sinh^{2}x-\sinh^{2}x'|,\label{scaling}
\end{equation}
where we have defined $s\equiv L/l$. This non-linear diffusion equation was
derived by Mello and Pichard \cite{Mel89} from a Fokker-Planck equation
\cite{Sto91,Dor82,Mel88} for the joint distribution function of all $N$
eigenvalues, by integrating out $N-1$ eigenvalues and taking the large-$N$
limit. This limit restricts its validity to the metallic regime ($N\gg L/l$),
and is sufficient to determine the leading order contribution to the average
conductance, which is ${\cal O}(N)$. The weak-localization correction, which is
${\cal O}(1)$, is neglected here. A priori, eq.\ (\ref{scaling}) holds only for
a ``quasi-one-dimensional'' wire geometry (length $L$ much greater than width
$W$), because the Fokker-Planck equation from which it is derived requires
$L\gg W$. Numerical simulations indicate that the geometry dependence only
appears in the ${\cal O}(1)$ corrections, and that the ${\cal O}(N)$
contributions are essentially the same for a wire, square, or cube.

In ref.\ \cite{Bee94b} it is shown how the scaling equation (\ref{scaling}) can
be solved exactly, for arbitrary initial condition
$\rho(x,0)\equiv\rho_{0}(x)$. The method of solution is based on a mapping of
eq.\ (\ref{scaling}) onto Euler's equation for the isobaric flow of a
two-dimensional ideal fluid: $L$ corresponds to time and $\rho$ to the
$y$-component of the velocity field on the $x$-axis. [Please note that in this
section $x$ is the auxiliary variable defined in eq.\ (\ref{Txdef}) and {\em
not\/} the physical coordinate in fig.\ \ref{diagram}.] The result is
\begin{equation}
\rho(x,s)=(2N/\pi)\,{\rm Im}\,U(x-{\rm i}0^{+},s),\label{rhoxU}
\end{equation}
where the complex function $U(z,s)$ is determined by
\begin{equation}
U(z,s)=U_{0}\bigl(z-sU(z,s)\bigr).\label{Udef}
\end{equation}
The function $U_{0}(z)$ is fixed by the initial condition,
\begin{equation}
U_{0}(z)=\frac{\sinh 2z}{2N} \int_{0}^{\infty}\!\!{\rm
d}x'\,\frac{\rho_{0}(x')}{\sinh^{2}z-\sinh^{2}x'}.\label{U0def}
\end{equation}
The implicit equation (\ref{Udef}) has multiple solutions in the entire complex
plane; We need the solution for which both $z$ and $z-sU(z,s)$ lie in the strip
between the lines $y=0$ and $y=-\pi/2$, where $z=x+{\rm i}y$.

\begin{figure}[tb]
\hspace*{\fill}
\psfig{figure=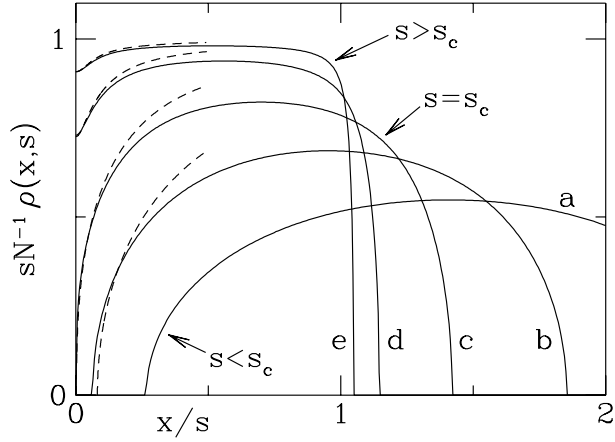,width=
8cm,bbllx=53pt,bblly=193pt,bburx=552pt,bbury=562pt}
\hspace*{\fill}
\medskip\caption[]{
Eigenvalue density $\rho(x,s)$ as a function of $x$ (in units of $s=L/l$) for
$\Gamma=0.1$. Curves a,b,c,d,e are for $s=2,4,9,30,100$, respectively. The
solid curves are from eq.\ (\protect\ref{rhoxU}), the dashed curves from eq.\
(\protect\ref{rhoxapprox}). The collision of the density profile with the
boundary at $x=0$, for $s=s_{\rm c}=(1-\Gamma)/\Gamma$, signals the
disorder-induced opening of tunneling channels responsible for the
reflectionless-tunneling effect. {\em (From ref.\ \protect\cite{Bee94b}.)}
\label{rhoplot}}
\end{figure}

The initial condition (\ref{rhox0}) corresponds to
\begin{equation}
U_{0}(z)={\textstyle\frac{1}{2}}\sinh 2z\,(\cosh^{2}z
-\Gamma^{-1})^{-1}.\label{U0x0}
\end{equation}
The resulting density (\ref{rhoxU}) is plotted in fig.\ \ref{rhoplot} (solid
curves), for $\Gamma=0.1$ and several values of $s$. For $s\gg 1$ and $x\ll s$
it simplifies to
\begin{eqnarray}
&&x={\textstyle\frac{1}{2}}{\rm arccosh}\,\tau-{\textstyle\frac{1}{2}}\Gamma
s(\tau^{2}-1)^{1/2}\cos\sigma,\nonumber\\
&&\sigma\equiv \pi sN^{-1}\rho(x,s),\;\;\tau\equiv\sigma(\Gamma
s\sin\sigma)^{-1},\label{rhoxapprox}
\end{eqnarray}
shown dashed in fig.\ \ref{rhoplot}. Equation (\ref{rhoxapprox}) agrees with
the result of a Green's function calculation by Nazarov \cite{Naz94}. For $s=0$
(no disorder), $\rho$ is a delta function at $x_{0}$. On adding disorder the
eigenvalue density rapidly spreads along the $x$-axis (curve a), such that
$\rho\leq N/s$ for $s>0$. The sharp edges of the density profile, so
uncharacteristic for a diffusion profile, reveal the hydrodynamic nature of the
scaling equation (\ref{scaling}). The upper edge is at
\begin{equation}
x_{\rm max}=s+{\textstyle\frac{1}{2}}\ln(s/\Gamma)+{\cal O}(1).\label{xmax}
\end{equation}
Since $L/x$ has the physical significance of a localization length
\cite{Sto91}, this upper edge corresponds to a minimum localization length
$\xi_{\rm min}=L/x_{\rm max}$ of order $l$. The lower edge at $x_{\rm min}$
propagates from $x_{0}$ to $0$ in a ``time'' $s_{\rm c}=(1-\Gamma)/\Gamma$. For
$1\ll s\leq s_{\rm c}$ one has
\begin{equation}
x_{\rm min}={\textstyle\frac{1}{2}}{\rm arccosh}\,(s_{\rm c}/s)
-{\textstyle\frac{1}{2}}[1-(s/s_{\rm c})^{2}]^{1/2}.\label{xmin}
\end{equation}
It follows that the maximum localization length $\xi_{\rm max}=L/x_{\rm min}$
{\em increases\/} if disorder is added to a tunnel junction. This paradoxical
result, that disorder enhances transmission, becomes intuitively obvious from
the hydrodynamic correspondence, which implies that $\rho(x,s)$ spreads both to
larger {\em and\/} smaller $x$ as the fictitious time $s$ progresses. When
$s=s_{\rm c}$ the diffusion profile hits the boundary at $x=0$ (curve c), so
that $x_{\rm min}=0$. This implies that for $s>s_{\rm c}$ there exist
scattering states (eigenfunctions of $tt^{\dagger}$) which tunnel through the
barrier with near-unit transmission probability, even if $\Gamma\ll 1$. The
number $N_{\rm open}$ of transmission eigenvalues close to one ({\em open
channels}) is of the order of the number of $x_{n}$'s in the range $0$ to $1$
(since $T_{n}\equiv 1/\cosh^{2}x_{n}$ vanishes exponentially if $x_{n}>1$). For
$s\gg s_{\rm c}$ (curve e) we estimate
\begin{equation}
N_{\rm open}\simeq\rho(0,s)=N(s+\Gamma^{-1})^{-1},\label{Nopen}
\end{equation}
where we have used eq.\ (\ref{rhoxapprox}). The disorder-induced opening of
tunneling channels was discovered by Nazarov \cite{Naz94}. It is the
fundamental mechanism for the $\Gamma^{-2}$ to $\Gamma^{-1}$ transition in the
conductance of an NS junction, as we now discuss.

According to eqs.\ (\ref{keyzero}), (\ref{Landauer}), (\ref{Txdef}), and
(\ref{rhoxdef}), the average conductances $\langle G_{\rm NS}\rangle$ and
$\langle G_{\rm N}\rangle$ are given by the integrals
\begin{eqnarray}
\langle G_{\rm NS}\rangle&=&\frac{4e^{2}}{h}\int_{0}^{\infty}\!
{\rm d}x\,\rho(x,s)\cosh^{-2}2x,\label{GNSx}\\
\langle G_{\rm N}\rangle&=&\frac{2e^{2}}{h}\int_{0}^{\infty}\!
{\rm d}x\,\rho(x,s)\cosh^{-2}x.\label{GNx}
\end{eqnarray}
Here we have used the same trigonometric identity as in eq.\ (\ref{GNSzeta}).
For $\Gamma\gg l/L$ one is in the regime $s\gg s_{\rm c}$ of curve e in fig.\
\ref{rhoplot}. Then the dominant contribution to the integrals comes from the
range $x/s\ll 1$ where $\rho(x,s)\approx\rho(0,s)=N(s+\Gamma^{-1})^{-1}$ is
approximately independent of $x$. Substitution of $\rho(x,s)$ by $\rho(0,s)$ in
eqs.\ (\ref{GNSx}) and (\ref{GNx}) yields directly
\begin{equation}
\langle G_{\rm NS}\rangle\approx\langle G_{\rm N}\rangle\approx 1/R_{\rm
N}^{\rm class},\label{GNSGN}
\end{equation}
in agreement with the result (\ref{RNSzeroB}) of the numerical simulations.

\begin{figure}[tb]
\hspace*{\fill}
\psfig{figure=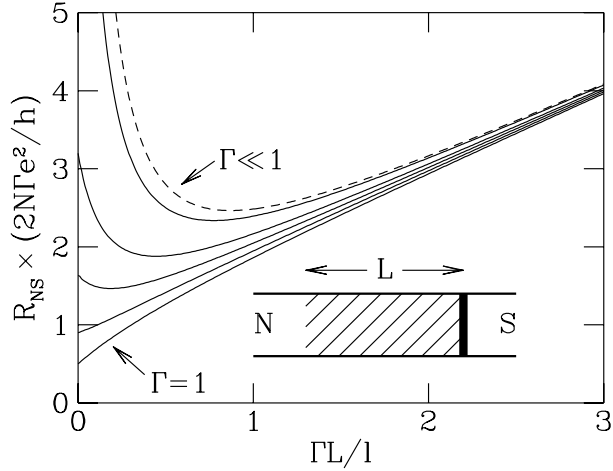,width=
8cm,bbllx=49pt,bblly=174pt,bburx=552pt,bbury=570pt}
\hspace*{\fill}
\medskip\caption[]{
Dependence of the resistance $R_{\rm NS}$ on the length $L$ of the disordered
normal region (hatched in the inset), for different values of the transmittance
$\Gamma$ of the NS interface. Solid curves are computed from eq.\
(\protect\ref{GNSresult}), for $\Gamma= 1,0.8,0.6,0.4,0.1$ from bottom to top.
For $\Gamma\ll 1$ the dashed curve is approached. {\em (From ref.\
\protect\cite{Bee94b}.)}
\label{GNSplot}}
\end{figure}

Equation (\ref{GNSGN}) has the linear $\Gamma$ dependence characteristic for
reflectionless tunneling. The crossover to the quadratic $\Gamma$ dependence
when $\Gamma\lesssim l/L$ is obtained by evaluating the integrals (\ref{GNSx})
and (\ref{GNx}) with the density $\rho(x,s)$ given by eq.\ (\ref{rhoxU}). The
result is \cite{Bee94b}
\begin{eqnarray}
\langle G_{\rm NS}\rangle&=&(2Ne^{2}/h)(s+Q^{-1})^{-1},\label{GNSresult}\\
\langle G_{\rm N}\rangle&=&(2Ne^{2}/h)(s+\Gamma^{-1})^{-1}.\label{GNresult}
\end{eqnarray}
The ``effective'' tunnel probability $Q$ is defined by
\begin{equation}
Q=\frac{\theta}{s\cos\theta}\left(\frac{\theta} {\Gamma
s\cos\theta}(1+\sin\theta)-1\right),\label{Qdef}
\end{equation}
where $\theta\in(0,\pi/2)$ is the solution of the transcendental equation
\begin{equation}
\theta[1-{\textstyle\frac{1}{2}}\Gamma(1-\sin\theta)]=\Gamma s\cos\theta.
\label{phidef}
\end{equation}
For $\Gamma\ll 1$ (or $s\gg 1$) eqs.\ (\ref{Qdef}) and (\ref{phidef}) simplify
to $Q=\Gamma\sin\theta$, $\theta=\Gamma s\cos\theta$, in precise agreement with
the Green's function calculation of Volkov, Za\u{\i}tsev, and Klapwijk
\cite{Vol93}. According to eq.\ (\ref{GNresult}), the normal-state resistance
increases {\em linearly\/} with the length $L$ of the disordered region, as
expected from Ohm's law. This classical reasoning fails if one of the contacts
is in the superconducting state. The scaling of the resistance $R_{\rm
NS}\equiv 1/\langle G_{\rm NS}\rangle$ with length, computed from eq.\
(\ref{GNSresult}), is plotted in fig.\ \ref{GNSplot}. For $\Gamma=1$ the
resistance increases monotonically with $L$. The ballistic limit $L\rightarrow
0$ equals $h/4Ne^{2}$, half the contact resistance of a normal junction because
of Andreev reflection (cf.\ section 3.1). For $\Gamma\lesssim 0.5$ a {\em
resistance minimum\/} develops, somewhat below $L=l/\Gamma$. The resistance
minimum is associated with the crossover from a quadratic to a linear
dependence of $R_{\rm NS}$ on $1/\Gamma$.

If $\Gamma s\gg 1$ one has $\theta\rightarrow\pi/2$, hence
$Q\rightarrow\Gamma$. In the opposite regime $\Gamma s\ll 1$ one has
$\theta\rightarrow\Gamma s$, hence $Q\rightarrow\Gamma^{2}s$. The corresponding
asymptotic expressions for $\langle G_{\rm NS}\rangle$ are (assuming $\Gamma\ll
1$ and $s\gg 1$):
\begin{eqnarray}
\langle G_{\rm NS}\rangle&=&(2Ne^{2}/h)(s+\Gamma^{-1})^{-1},\;\;{\rm if}\;\;
\Gamma s\gg 1,\label{asympta}\\
\langle G_{\rm NS}\rangle&=&(2Ne^{2}/h)\Gamma^{2}s,\;\;{\rm if}\;\; \Gamma s\ll
1.\label{asymptb}
\end{eqnarray}
In either limit the conductance is greater than the classical result
\begin{equation}
G_{\rm NS}^{\rm class}=(2Ne^{2}/h)(s+2\Gamma^{-2})^{-1}, \label{GNSclass}
\end{equation}
which holds if phase coherence between electrons and holes is destroyed by a
voltage or magnetic field. The peak in the conductance around $V,B=0$ is of
order $\Delta G_{\rm NS}=\langle G_{\rm NS}\rangle-G_{\rm NS}^{\rm class}$,
which has the relative magnitude
\begin{equation}
\frac{\Delta G_{\rm NS}}{\langle G_{\rm
NS}\rangle}\approx\frac{2}{2+\Gamma^{2}s}.\label{peakheight}
\end{equation}

The scaling theory assumes zero temperature. Hekking and Nazarov \cite{Hek93}
have studied the conductance of a resistive NS interface at finite
temperatures, when $L$ is greater than the correlation length $L_{\rm c}={\rm
min}\,(l_{\phi},\sqrt{\hbar D/k_{\rm B}T})$. Their result is consistent with
the limiting expression (\ref{asymptb}), if $s=L/l$ is replaced by $L_{\rm
c}/l$. The implication is that, if $L>L_{\rm c}$, the non-linear scaling of the
resistance shown in fig.\ \ref{GNSplot} only applies to a disordered segment of
length $L_{\rm c}$ adjacent to the superconductor. For the total resistance one
should add the Ohmic contribution of order $(h/e^{2})(L-L_{\rm c})/l$ from the
rest of the wire.

\subsection{Double-barrier junction}

In the previous subsection we have discussed how the opening of tunneling
channels (i.e.\ the appearance of transmission eigenvalues close to one) by
disorder leads to a minimum in the resistance when $L\simeq l/\Gamma$. The
minimum separates a $\Gamma^{-1}$ from a $\Gamma^{-2}$ dependence of the
resistance on the transparency of the interface. We referred to the
$\Gamma^{-1}$ dependence as ``reflectionless tunneling'', since it is as if one
of the two quasiparticles which form the Cooper pair can tunnel through the
barrier with probability one. In the present subsection we will show, following
ref.\ \cite{Mel94}, that a qualitatively similar effect occurs if the disorder
in the normal region is replaced by a second tunnel barrier (tunnel probability
$\Gamma'$). The resistance at fixed $\Gamma$ shows a minimum as a function of
$\Gamma'$ when $\Gamma'\simeq\Gamma$.  For $\Gamma'\lesssim\Gamma$ the
resistance has a $\Gamma^{-1}$ dependence, so that we can speak again of
reflectionless tunneling.

We consider an ${\rm NI}_{1}{\rm NI}_{2}{\rm S}$ junction, where N = normal
metal, S = superconductor, and ${\rm I}_i$ = insulator or tunnel barrier
(transmission probability per mode $\Gamma_{i}\equiv 1/\cosh^{2}\alpha_{i}$).
We assume ballistic motion between the barriers. (The effect of disorder is
discussed later.) A straightforward calculation yields the transmission
probabilities $T_{n}$ of the two barriers in series,
\begin{eqnarray}
&&T_{n}=(a+b\cos\varphi_{n})^{-1},\label{eq:tnphin}\\
&&a={\textstyle\frac{1}{2}}+{\textstyle\frac{1}{2}}\cosh 2\alpha_{1}\cosh
2\alpha_{2},\;\;
b={\textstyle\frac{1}{2}}\sinh 2\alpha_{1}\sinh 2\alpha_{2},\label{coeffalpha}
\end{eqnarray}
where $\varphi_{n}$ is the phase accumulated between the barriers by mode $n$.
Since the transmission matrix $t$ is diagonal, the transmission probabilities
$T_n$ are identical to the eigenvalues of $tt^{\dagger}$. We assume that
$L\gg\lambda_{\rm F}$ ($\lambda_{\rm F}$ is the Fermi wavelength) and
$N\Gamma_{i}\gg 1$, so that the conductance is not dominated by a single
resonance. In this case, the phases $\varphi_{n}$ are distributed uniformly in
the interval $(0,2\pi)$ and we may replace the sum over the transmission
eigenvalues in eqs.\ (\ref{keyzero}) and (\ref{Landauer}) by integrals over
$\varphi$: $\sum_{n=1}^{N}f(\varphi_{n})\rightarrow(N/2\pi)\int_{0}^{2\pi}{\rm
d}\varphi\,f(\varphi)$. The result is
\begin{eqnarray}
G_{\rm NS}&=&\frac{4Ne^2}{h}\frac{\cosh 2\alpha_{1}\cosh 2\alpha_{2}} {\left(
\cosh^{2}2\alpha_{1}+\cosh^{2}2\alpha_{2}-1 \right)^{3/2}},\label{gnsintphi}\\
G_{\rm N}&=&\frac{4Ne^{2}}{h}(\cosh 2\alpha_{1}+\cosh
2\alpha_{2})^{-1}\label{gnintphi}.
\end{eqnarray}
These expressions are symmetric in the indices 1 and 2: It does not matter
which of the two barriers is closest to the superconductor. In the same way we
can compute the entire distribution of the transmission eigenvalues,
$\rho(T)\equiv\sum_{n}\delta(T-T_{n}) \rightarrow(N/2\pi)\int_{0}^{2\pi}{\rm
d}\varphi\,\delta(T-T(\varphi))$. Substituting
$T(\varphi)=(a+b\cos\varphi)^{-1}$ from eq.\ (\ref{eq:tnphin}), one finds
\begin{equation}
\rho(T)=\frac{N}{\pi T}\left(b^{2}T^{2}-(aT-1)^{2}\right)^{-1/2}.
\label{rhoT}
\end{equation}

\begin{figure}[tb]
\hspace*{\fill}
\psfig{figure=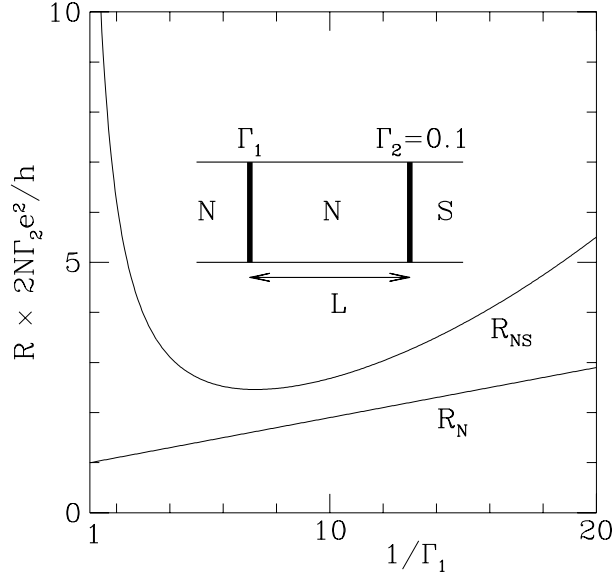,width=
8cm,bbllx=19pt,bblly=160pt,bburx=541pt,bbury=661pt}
\hspace*{\fill}
\medskip\caption[]{Dependence of the resistances $R_{\rm N}$ and $R_{\rm NS}$ of
ballistic NININ and NINIS structures, respectively, on barrier transparency
$\Gamma_{1}$, while transparency $\Gamma_{2}=0.1$ is kept fixed [computed from
eqs.\ (\protect\ref{gnsintphi}) and (\protect\ref{gnintphi})]. The inset shows
the NINIS structure considered. {\em (From ref.\ \protect\cite{Mel94}.)}
\label{NINISfig}}
\end{figure}

In fig.\ \ref{NINISfig} we plot the resistance $R_{\rm N}= 1/G_{\rm N}$ and
$R_{\rm NS}=1/G_{\rm NS}$, following from eqs.\ (\ref{gnsintphi}) and
(\ref{gnintphi}). Notice that $R_{\rm N}$ follows Ohm's law,
\begin{equation}
R_{\rm N}=\frac{h}{2Ne^2}(1/\Gamma_{1}+1/\Gamma_{2}-1),\label{Ohmslaw}
\end{equation}
as expected from classical considerations. In contrast, the resistance $R_{\rm
NS}$ has a {\em minimum\/} if one of the $\Gamma$'s is varied while keeping the
other fixed. This resistance minimum cannot be explained by classical series
addition of barrier resistances. If $\Gamma_{2}\ll 1$ is fixed and $\Gamma_{1}$
is varied, as in fig.\ \ref{NINISfig}, the minimum
occurs when $\Gamma_{1}=\sqrt{2}\,\Gamma_{2}$. The minimal resistance $R_{\rm
NS}^{\rm min}$ is of the same order of magnitude as the resistance $R_{\rm N}$
in the normal state at the same value of $\Gamma_{1}$ and $\Gamma_{2}$. In
particular, we find that $R_{\rm NS}^{\rm min}$ depends linearly on
$1/\Gamma_{i}$, whereas for a single barrier $R_{\rm NS}\propto 1/\Gamma^{2}$.

\begin{figure}[tb]
\hspace*{\fill}
\psfig{figure=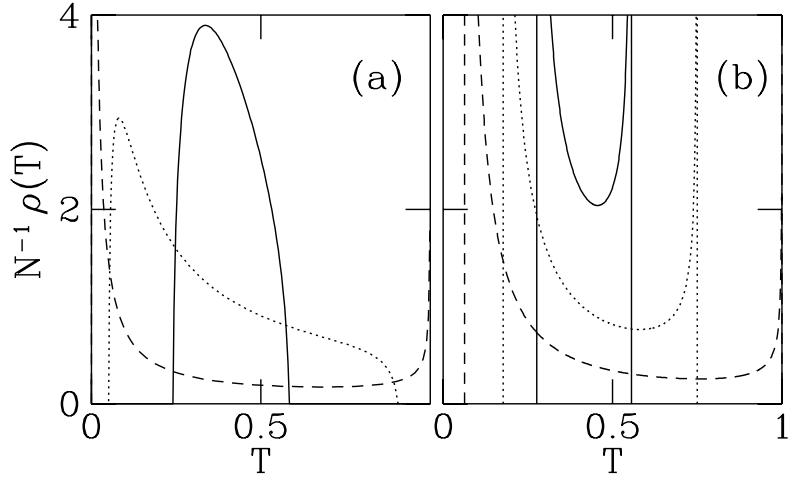,width=
10cm,bbllx=76pt,bblly=183pt,bburx=583pt,bbury=504pt}
\hspace*{\fill}
\medskip\caption[]{Density of transmission eigenvalues through a normal region
containing a potential barrier (transmission probability $\Gamma=0.4$). The
left panel (a) shows the disorder-induced opening of tunneling channels (solid
curve: $s=0.04$; dotted: $s=0.4$; dashed: $s=5$; where $s\equiv L/l$). The
right panel (b) shows the opening of channels by a second tunnel barrier
(transparency $\Gamma'$; solid curve: $\Gamma'=0.95$; dotted: $\Gamma'=0.8$;
dashed: $\Gamma'=0.4$). The curves in (a) are computed from eq.\
(\protect\ref{rhoxU}), the curves in (b) from eq.\ (\protect\ref{rhoT}). {\em
(From ref.\ \protect\cite{Mel94}.)}
\label{rhotplot}}
\end{figure}

The linear dependence on the barrier transparency shows the qualitative
similarity of a ballistic NINIS junction to the disordered NIS junction
considered in the previous subsection. To illustrate the similarity, we compare
in fig.\ \ref{rhotplot} the densities of normal-state transmission eigenvalues.
The left panel is for an NIS junction [computed using eq.\ (\ref{rhoxU})], the
right panel is for an NINIS junction [computed from eq.\ (\ref{rhoT})]. In the
NIS junction, disorder leads to a bimodal distribution $\rho(T)$, with a peak
near zero transmission and another peak near unit transmisssion (dashed curve).
A similar bimodal distribution appears in the ballistic NINIS junction, for
approximately equal transmission probabilities of the two barriers. There are
also differences between the two cases: The NIS junction has a uni-modal
$\rho(T)$ if $L/l<1/\Gamma$, while the NINIS junction has a bimodal $\rho(T)$
for any ratio of $\Gamma_{1}$ and $\Gamma_{2}$. In both cases, the opening of
tunneling channels, i.e.\ the appearance of a peak in $\rho(T)$ near $T=1$, is
the origin for the $1/\Gamma$ dependence of the resistance.

The scaling equation of section 5.2 can be used to investigate what happens to
the resistance minimum if the region of length $L$ between the tunnel barriers
contains impurities, with elastic mean free path $l$. As shown in ref.\
\cite{Mel94}, the resistance minimum persists as long as $l\gtrsim\Gamma L$.
In the diffusive regime ($l\ll L$) the scaling theory is found to agree with
the Green's function calculation by Volkov, Za\u{\i}tsev, and Klapwijk for a
disordered NINIS junction \cite{Vol93}. For strong barriers
($\Gamma_{1},\Gamma_{2}\ll 1$) and strong disorder ($L\gg l$), one has the two
asymptotic formulas
\begin{eqnarray}
G_{\rm NS}&=& \frac{2Ne^2}{h}\frac{\Gamma_{1}^{2}\Gamma_{2}^{2}}
{\left(\Gamma_{1}^{2}+\Gamma_{2}^{2}\right)^{3/2}},\;\;{\rm if}\;\;
\Gamma_{1},\Gamma_{2}\ll l/L,\label{eq:glimitls}\\
G_{\rm NS}&=&\frac{2Ne^2}{h}(L/l+1/\Gamma_{1}+1/\Gamma_{2})^{-1},\;\;{\rm
if}\;\;\Gamma_{1},\Gamma_{2}\gg l/L.\label{eq:glimitss}
\end{eqnarray}
Equation (\ref{eq:glimitls}) coincides with eq.\ (\ref{gnsintphi}) in the limit
$\alpha_{1},\alpha_{2}\gg 1$ (recall that $\Gamma_{i}\equiv
1/\cosh^2\alpha_{i}$). This shows that the effect of disorder on the resistance
minimum can be neglected as long as the resistance of the junction is dominated
by the barriers. In this case $G_{\rm NS}$ depends linearly on $\Gamma_{1}$ and
$\Gamma_{2}$ only if $\Gamma_{1}\approx\Gamma_{2}$. Equation
(\ref{eq:glimitss}) shows that if the disorder dominates, $G_{\rm NS}$ has a
linear $\Gamma$-dependence regardless of the relative magnitude of $\Gamma_{1}$
and $\Gamma_{2}$.

We have assumed zero temperature, zero magnetic field, and infinitesimal
applied voltage. Each of these quantities is capable of destroying the phase
coherence between the electrons and the Andreev-reflected holes, which is
responsible for the resistance minimum. As far as the temperature $T$ and
voltage $V$ are concerned, we require $k_{\rm B}T,eV\ll\hbar/\tau_{\rm dwell}$
for the appearance of a resistance minimum, where $\tau_{\rm dwell}$ is the
dwell time of an electron in the region between the two barriers.
For a ballistic NINIS junction $\tau_{\rm dwell}\simeq L/v_{\rm F}\Gamma$,
while for a disordered junction $\tau_{\rm dwell}\simeq L^{2}/v_{\rm F}\Gamma
l$ is larger by a factor $L/l$. It follows that the condition on temperature
and voltage becomes more restrictive if the disorder increases, even if the
resistance remains dominated by the barriers. As far as the magnetic field $B$
is concerned, we require $B\ll h/eS$ (with $S$ the area of the junction
perpendicular to $B$), if the motion between the barriers is diffusive. For
ballistic motion the trajectories enclose no flux, so no magnetic field
dependence is expected.

A possible experiment to verify these results might be scanning tunneling
microscopy (STM) of a metal particle on a superconducting substrate
\cite{Hesl94}. The metal--superconductor interface has a fixed tunnel
probability $\Gamma_{2}$. The probability $\Gamma_{1}$ for an electron to
tunnel from STM to particle can be controlled by varying the distance. (Volkov
has recently analyzed this geometry in the regime that the motion from STM to
particle is diffusive rather than by tunneling \cite{Vol94}.)
Another possibility is to create an NINIS junction using a two-dimensional
electron gas in contact with a superconductor. An adjustable tunnel barrier
could then be implemented by means of a gate electrode.

\subsection{Circuit theory}

The scaling theory of ref.\ \cite{Bee94b}, which was the subject of section
5.2, describes the transition from the ballistic to the diffusive regime. In
the diffusive regime it is equivalent to the Green's function theory of ref.\
\cite{Vol93}. A third, equivalent, theory for the diffusive regime was
presented recently by Nazarov \cite{Naz94b}. Starting from a continuity
equation for the Keldysh Green's function \cite{Lar77}, and applying the
appropriate boundary conditions \cite{Kup88}, Nazarov was able to formulate a
set of rules which reduce the problem of computing the resistance of an NS
junction to a simple exercise in circuit theory. Furthermore, the approach can
be applied without further complications to multi-terminal networks involving
several normal and superconducting reservoirs. Because of its practical
importance, we discuss Nazarov's circuit theory in some detail.

The superconductors $S_{i}$ should all be at the same voltage, but may have a
different phase $\phi_{i}$ of the pair potential. Zero temperature is assumed,
as well as infinitesimal voltage differences between the normal reservoirs
(linear response). The reservoirs are connected by a set of diffusive
normal-state conductors (length $L_{i}$, mean free path $l_{i}$; $s_{i}\equiv
L_{i}/l_{i}\gg 1$). Between the conductors there may be tunnel barriers (tunnel
probability $\Gamma_{i}$). The presence of superconducting reservoirs has no
effect on the resistance $(h/2Ne^{2})s_{i}$ of the diffusive conductors, but
affects only the resistance $h/2Ne^{2}\Gamma_{i}^{\rm eff}$ of the tunnel
barriers. The tunnel probability $\Gamma_{i}$ of barrier $i$ is renormalized to
an effective tunnel probability $\Gamma_{i}^{\rm eff}$, which depends on the
entire circuit.

Nazarov's rules to compute the effective tunnel probabilities are as follows.
To each node and to each terminal of the circuit one assigns a vector
$\vec{n}_{i}$ of unit length. For a normal reservoir, $\vec{n}_{i}=(0,0,1)$ is
at the north pole, for a superconducting reservoir,
$\vec{n}_{i}=(\cos\phi_{i},\sin\phi_{i},0)$ is at the equator. For a node,
$\vec{n}_{i}$ is somewhere on the northern hemisphere. The vector $\vec{n}_{i}$
is called a ``spectral vector'', because it is a particular parameterization of
the local energy spectrum. If the tunnel barrier is located between spectral
vectors $\vec{n}_{1}$ and $\vec{n}_{2}$, its effective tunnel probability
is\footnote{
It may happen that $\cos\theta_{12}<0$, in which case the effective tunnel
probability is negative. Nazarov has given an example of a four-terminal
circuit with $\Gamma^{\rm eff}<0$, so that the current through this barrier
flows in the direction opposite to the voltage drop \cite{Naz94c}.}
\begin{equation}
\Gamma^{\rm eff}=(\vec{n}_{1}\cdot\vec{n}_{2})\Gamma=
\Gamma\cos\theta_{12},\label{Gamma-eff}
\end{equation}
where $\theta_{12}$ is the angle between $\vec{n}_{1}$ and $\vec{n}_{2}$. The
rule to compute the spectral vector of node $i$ follows from the continuity
equation for the Green's function. Let the index $k$ label the nodes or
terminals connected to node $i$ by a single tunnel barrier (with tunnel
probability $\Gamma_{k}$). Let the index $q$ label the nodes or terminals
connected to $i$ by a diffusive conductor (with $L/l\equiv s_{q}$). The
spectral vectors then satisfy the sum rule \cite{Naz94b}
\begin{equation}
\sum_{k}(\vec{n}_{i}\times\vec{n}_{k})\Gamma_{k}+
\sum_{q}(\vec{n}_{i}\times\vec{n}_{q})\frac{{\rm
arccos}(\vec{n}_{i}\cdot\vec{n}_{q})}
{s_{q}\sqrt{1-(\vec{n}_{i}\cdot\vec{n}_{q})^{2}}}=0.\label{sumrule}
\end{equation}
This is a sum rule for a set of vectors perpendicular to $\vec{n}_{i}$ of
magnitude $\Gamma_{k}\sin\theta_{ik}$ or $\theta_{iq}/s_{q}$, depending on
whether the element connected to node $i$ is a tunnel barrier or a diffusive
conductor. There is a sum rule for each node, and together the sum rules
determine the spectral vectors of the nodes.

\begin{figure}[tb]
\hspace*{\fill}
\psfig{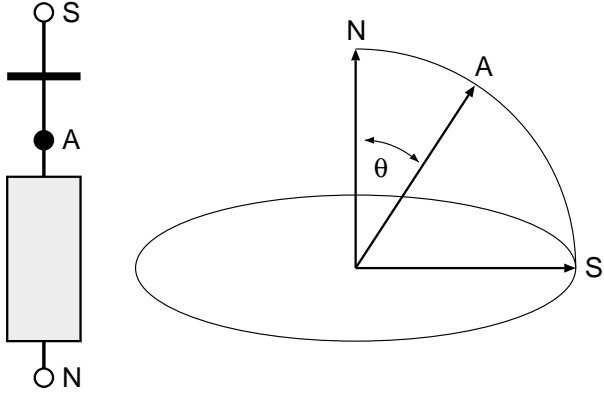}
\hspace*{\fill}
\medskip\caption[]{
At left: Circuit containing two terminals (open circles), one node (filled
circle), and two elements: A diffusive conductor (shaded) and a tunnel barrier
(black). At right: Spectral vectors associated with the terminals N,S and with
the node A.
\label{NScircuit}}
\end{figure}

As a simple example, let us consider the system of section 5.2, consisting of
one normal terminal (N), one superconducting terminal (S), one node (labeled
A), and two elements: A diffusive conductor (with $L/l\equiv s$) between N and
A, and a tunnel barrier (tunnel probability $\Gamma$) between A and S (see
fig.\ \ref{NScircuit}). There are three spectral vectors, $\vec{n}_{\rm N}$,
$\vec{n}_{\rm S}$, and $\vec{n}_{\rm A}$. All spectral vectors lie in one
plane. (This holds for any network with a single superconducting terminal.) The
resistance of the circuit is given by $R=(h/2Ne^{2})(s+1/\Gamma^{\rm eff})$,
with the effective tunnel probability
\begin{equation}
\Gamma^{\rm eff}=\Gamma\cos\theta_{\rm AS}=\Gamma\sin\theta.\label{thetaAS}
\end{equation}
Here $\theta\in[0,\pi/2]$ is the polar angle of $\vec{n}_{\rm A}$. This angle
is determined by the sum rule (\ref{sumrule}), which in this case takes the
form
\begin{equation}
\Gamma\cos\theta-\theta/s=0.\label{sumruleAS}
\end{equation}
Comparison with section 5.2 shows that $\Gamma^{\rm eff}$ coincides with the
effective tunnel probability $Q$ of eq.\ (\ref{Qdef}) in the limit $s\gg 1$,
i.e.\ if one restricts oneself to the diffusive regime. That is the basic
requirement for the application of the circuit theory.

\begin{figure}[tb]
\hspace*{\fill}
\psfig{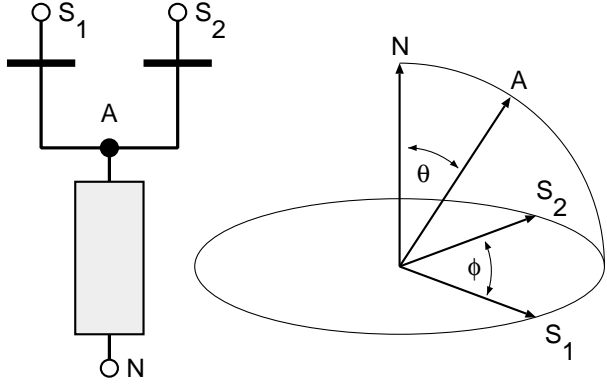}
\hspace*{\fill}
\medskip\caption[]{
Circuit diagram and spectral vectors for a structure containing one normal and
two superconducting terminals (phase difference $\phi$).
\label{forkcircuit}}
\end{figure}

Let us now consider the ``fork junction'' of fig.\ \ref{forkcircuit}, with one
normal terminal (N) and two superconducting terminals ${\rm S}_{1}$ and ${\rm
S}_{2}$ (phases $\phi_{1}\equiv-\phi/2$ and $\phi_{2}\equiv\phi/2$). There is
one node (A), which is connected to N by a diffusive conductor ($L/l\equiv s$),
and to ${\rm S}_{1}$ and ${\rm S}_{2}$ by tunnel barriers ($\Gamma_{1}$ and
$\Gamma_{2}$). This structure was studied theoretically by Hekking and Nazarov
\cite{Hek93} and experimentally by Pothier et al.\ \cite{Pot94}. For
simplicity, let us assume two identical tunnel barriers
$\Gamma_{1}=\Gamma_{2}\equiv\Gamma$. Then the spectral vector $\vec{n}_{\rm
A}=(\sin\theta,0,\cos\theta)$ of node A lies symmetrically between the spectral
vectors of terminals ${\rm S}_{1}$ and ${\rm S}_{2}$. The sum rule
(\ref{sumrule}) now takes the form
\begin{equation}
2\Gamma|\cos{\textstyle\frac{1}{2}}\phi|
\cos\theta-\theta/s=0.\label{sumrulefork}
\end{equation}
Its solution determines the effective tunnel rate $\Gamma^{\rm
eff}=\Gamma|\cos{\textstyle\frac{1}{2}}\phi|\sin\theta$ of each of the two
barriers in parallel, and hence the conductance of the fork junction,
\begin{equation}
G=\frac{2Ne^{2}}{h}[s+{\textstyle\frac{1}{2}}(\Gamma
|\cos{\textstyle\frac{1}{2}}\phi|\sin\theta)^{-1}]^{-1}.\label{Gfork}
\end{equation}
Two limiting cases of eqs.\ (\ref{sumrulefork}) and (\ref{Gfork}) are
\begin{eqnarray}
G&=&(2Ne^{2}/h)(s+{\textstyle\frac{1}{2}}\Gamma^{-1}
|\cos{\textstyle\frac{1}{2}}\phi|^{-1})^{-1},\;\;{\rm
if}\;\;s\Gamma|\cos{\textstyle\frac{1}{2}}\phi|\gg 1,\label{Gforka}\\
G&=&(4Ne^{2}/h)s\Gamma^{2}(1+\cos\phi),\;\;{\rm
if}\;\;s\Gamma|\cos{\textstyle\frac{1}{2}}\phi|\ll 1.\label{Gforkb}
\end{eqnarray}
For $\phi=0$ (and $2\Gamma\rightarrow\Gamma$) these expressions reduce to the
results (\ref{asympta}) and (\ref{asymptb}) for an NS junction with a single
superconducting reservoir. The limit (\ref{Gforkb}) agrees with the
finite-temperature result of Hekking and Nazarov \cite{Hek93}, if $s$ is
replaced by $L_{\rm c}/l$ and a series resistance is added due to the normal
segment which is further than a correlation length from the NS interfaces. The
possibility of a dependence of the conductance on the superconducting phase
difference was noted also in other theoretical works, for different geometries
\cite{Spi82,Alt87,Nak91,Tak92,Lam93,Hui93}.

\begin{figure}[tb]
\hspace*{\fill}
\psfig{figure=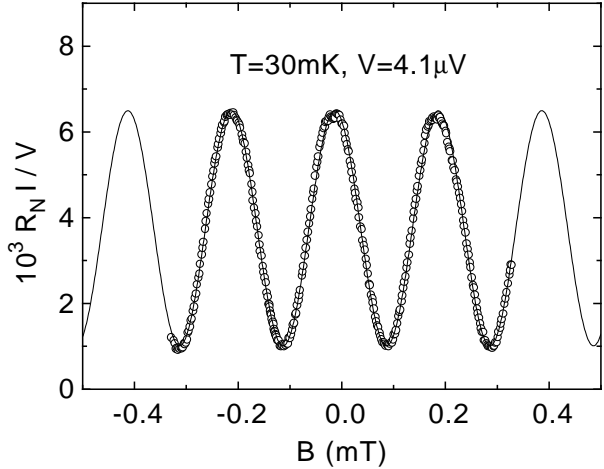,width=
8cm,bbllx=188pt,bblly=445pt,bburx=559pt,bbury=736pt}
\hspace*{\fill}
\medskip\caption[]{Conductance of a fork junction as a function of magnetic field,
showing the dependence on the phase difference $\phi$ of the superconductor at
two tunnel barriers. The circles are measurements by Pothier et al.\
\protect\cite{Pot94} of the current $I$ through a Cu wire connected to an
oxidized Al fork (normal-state resistance $R_{\rm N}=1.56\,{\rm k}\Omega$). The
applied voltage $V$ is sufficiently low that $I/V$ is close to the
linear-response conductance. (The amplitude of the oscillations at $V=0$ is
$3.94\cdot 10^{-6}\,\Omega^{-1}$, somewhat larger than in the figure.) The
solid curve is a cosine fit to the data. The offset of maximum conductance from
$B=0$ is attributed to a small residual field in the cryostat. {\em (Courtesy
of H. Pothier.)}
\label{Pothier}}
\end{figure}

The $\phi$-dependence of the conductance of a fork junction has recently been
observed by Pothier et al.\ \cite{Pot94}. Some of their data is reproduced in
fig.\ \ref{Pothier}. The conductance of a Cu wire attached to an oxidized Al
fork oscillates as a function of the applied magnetic field. The period
corresponds to a flux increment of $h/2e$ through the area enclosed by the fork
and the wire, and thus to $\Delta\phi=2\pi$. The experiment is in the regime
where the junction resistance is dominated by the tunnel barriers, as in eq.\
(\ref{Gforkb}).\footnote{
Equation (\protect\ref{Gforkb}) provides only a qualitative description of the
experiment, mainly because the motion in the arms of the fork is diffusive
rather than ballistic. This is why the conductance minima in fig.\
\protect\ref{Pothier} do not go to zero. A solution of the diffusion equation
in the actual experimental geometry is required for a quantitative comparison
with the theory \protect\cite{Pot94}.}
The metal-oxide tunnel barriers in such structures have typically very small
transmission probabilities ($\Gamma\simeq 10^{-5}$ in ref.\ \cite{Pot94}), so
that the regime of eq.\ (\ref{Gforka}) is not easily accessible. Larger
$\Gamma$'s can be realized by the Schottky barrier at a semiconductor ---
superconductor interface. It would be of interest to observe the crossover with
increasing $\Gamma$ to the non-sinusoidal $\phi$-dependence predicted by eq.\
(\ref{Gforka}), as a further test of the theory.

\section{Universal conductance fluctuations}

\begin{figure}[tb]
\hspace*{\fill}
\psfig{figure=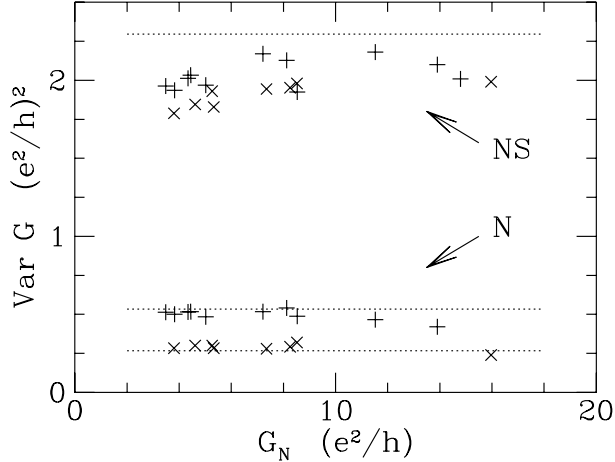,width=
8cm,bbllx=53pt,bblly=171pt,bburx=560pt,bbury=562pt}
\hspace*{\fill}
\medskip\caption[]{Numerical calculation of the variance of the fluctuations in $G_{\rm
N}$ and $G_{\rm NS}$, as a function of the average $G_{\rm N}$ ($+$ for $B=0$;
$\times$ for a flux of $10\,h/e$). Dotted lines are the analytical results from
eqs.\ (\protect\ref{UCFN}) and (\protect\ref{UCFNS}). Note the absence of a
factor-of-two reduction in ${\rm Var}\,G_{\rm NS}$ on applying a magnetic
field. {\em (From ref.\ \protect\cite{Mar93}.)}
\label{variance}}
\end{figure}

So far we have considered the {\em average\/} of the conductance over an
ensemble of impurity potentials. In fig.\ \ref{variance} we show results of
numerical simulations \cite{Mar93} for the {\em variance\/} of the
sample-to-sample fluctuations of the conductance, as a function of the average
conductance in the normal state. A range of parameters $L,W,l,N$ was used to
collect this data, in the quasi-one-dimensional, metallic, diffusive regime
$l<W<L<Nl$. An ideal NS interface was assumed ($\Gamma=1$). The results for
${\rm Var\,}G_{\rm N}$ are as expected theoretically \cite{Sto91,Mel91} for
``universal conductance fluctuations'' (UCF):
\begin{equation}
{\rm Var\,}G_{\rm N}=\frac{8}{15}\,\beta^{-1}(e^{2}/h)^{2}.\label{UCFN}
\end{equation}
The index $\beta$ equals 1 in the presence and 2 in the absence of
time-reversal symmetry. The $1/\beta$ dependence of ${\rm Var\,}G_{\rm N}$
implies that the variance of the conductance fluctuations is reduced by a
factor of two upon application of a magnetic field, as observed in the
simulation (see the two dotted lines in the lower part of fig.\
\ref{variance}). The data for ${\rm Var\,}G_{\rm NS}$ at $B=0$ shows
approximately a four-fold increase over ${\rm Var\,}G_{\rm N}$. For $B\neq 0$,
the simulation shows that ${\rm Var\,}G_{\rm NS}$ is essentially {\em
unaffected\/} by a time-reversal-symmetry breaking magnetic field. In contrast
to the situation in the normal state, the theory for UCF in an NS junction is
quite different for zero and for non-zero magnetic field, as we now discuss.

In zero magnetic field, the conductance of the NS junction is given by eq.\
(\ref{keyzero}), which is an expression of the form $A=\sum_{n}a(T_{n})$. Such
a quantity $A$ is called a {\em linear statistic\/} on the transmission
eigenvalues. The word ``linear'' refers to the fact that $A$ does not contain
products of different $T_{n}$'s. The function $a(T)$ may well depend
non-linearly on $T$, as it does for $G_{\rm NS}$, where $a(T)$ is a rational
function of $T$. The Landauer formula (\ref{Landauer}) for the normal-state
conductance is also a linear statistic, with $a(T)\propto T$. It is a general
theorem in random-matrix theory \cite{Bee93} that the variance of a linear
statistic has a $1/\beta$ dependence on the symmetry index $\beta$. Moreover,
the magnitude of the variance is independent of the microscopic properties of
the system (sample size, degree of disorder). This is Imry's fundamental
explanation for UCF \cite{Imr86}.

For a wire geometry, there exists a formula for the variance of an arbitrary
linear statistic \cite{Mac94,Bee93b,Cha94},
\begin{eqnarray}
{\rm Var}\,A=-\frac{1}{2\beta\pi^{2}}\int_{0}^{1}
\!\!dT\int_{0}^{1}
\!\!dT'\left(\frac{da(T)}{dT}\right)
\left(\frac{da(T')}{dT'}\right)\nonumber\\
\times\ln\left(\frac{1+\pi^{2}[x(T)+x(T')]^{-2}}
{1+\pi^{2}[x(T)-x(T')]^{-2}}\right),
\label{VarAresult}
\end{eqnarray}
where $x(T)={\rm arccosh}\,T^{-1/2}$. In the normal state, substitution of
$a(T)=(2e^{2}/h)T$ into eq.\ (\ref{VarAresult}) reproduces the result
(\ref{UCFN}). In the NS junction, substitution of
$a(T)=(4e^{2}/h)T^{2}(2-T)^{-2}$ yields, for the case $\beta=1$ of zero
magnetic field,
\begin{equation}
{\rm Var\,}G_{\rm NS}=\frac{32}{15}(2-90\pi^{-4})(e^{2}/h)^{2}=4.30\,{\rm
Var}\,G_{\rm N}.\label{UCFNS}
\end{equation}
A factor of four between ${\rm Var\,}G_{\rm NS}$ and ${\rm Var\,}G_{\rm N}$ was
estimated by Takane and Ebisawa \cite{Tak92b}, by an argument similar to that
which we described in section 4 for the weak-localization correction. (A
diagrammatic calculation by the same authors \cite{Tak91} gave a factor of six,
presumably because only the dominant diagram was included.) The numerical data
in fig.\ \ref{variance} is within 10~\% of the theoretical prediction
(\ref{UCFNS}) (upper dotted line). Similar numerical results for ${\rm
Var\,}G_{\rm NS}$ in zero magnetic field were obtained in refs.\
\cite{Tak92b,Bru94}.

We conclude that UCF in zero magnetic field is basically the same phenomenon
for $G_{\rm N}$ and $G_{\rm NS}$, because both quantities are linear statistics
for $\beta=1$. If time-reversal symmetry (TRS) is broken by a magnetic field,
the situation is qualitatively different. For $G_{\rm N}$, broken TRS does not
affect the universality of the fluctuations, but merely reduces the variance by
a factor of two. No such simple behavior is to be expected for $G_{\rm NS}$,
since it is no longer a linear statistic for $\beta=2$. That is a crucial
distinction between eq.\ (\ref{key}) for $G_{\rm NS}$ and the Landauer formula
(\ref{Landauer}) for $G_{\rm N}$, which remains a linear statistic regardless
of whether TRS is broken or not. This expectation \cite{Bee92} of an anomalous
$\beta$-dependence of ${\rm Var}\,G_{\rm NS}$ was borne out by numerical
simulations \cite{Mar93}, which showed that the conductance fluctuations in an
NS junction without TRS remain independent of disorder, and of approximately
the same magnitude as in the presence of TRS (compare $+$ and $\times$ data
points in the upper part of fig.\ \ref{variance}). An analytical theory remains
to be developed.

\section{Shot noise}

The conductance, which we studied in the previous sections, is the {\em
time-averaged\/} current $I$ divided by the applied voltage $V$. Time-dependent
fluctuations $\delta I(t)$ in the current give additional information on the
transport processes. The zero-frequency noise power $P$ is defined by
\begin{equation}
P=4\int_{0}^{\infty}{\rm d}t\,\langle\delta I(t)\delta I(0)\rangle.\label{Pdef}
\end{equation}
At zero temperature, the discreteness of the electron charge is the only source
of fluctuations in time of the current. These fluctuations are known as ``shot
noise'', to distinguish them from the thermal noise at non-zero temperature. A
further distinction between the two is that the shot-noise power is
proportional to the applied voltage, whereas the thermal noise does not vanish
at $V=0$. Shot noise is therefore an intrinsically non-equilibrium phenomenon.
If the transmission of an elementary charge $e$ can be regarded as a sequence
of uncorrelated events, then $P=2e|I|\equiv P_{\rm Poisson}$ as in a Poisson
process. In this section we discuss, following ref.\ \cite{Jon94}, the
enhancement of shot noise in an NS junction. The enhancement originates from
the fact that the current in the superconductor is carried by Cooper pairs in
units of $2e$. However, as we will see, a simple factor-of-two enhancement
applies only in certain limiting cases.

In the normal state, the shot-noise power (at zero temperature and
infinitesimal applied voltage) is given by \cite{But90b}
\begin{equation}
P_{\rm N}=P_{0}{\rm Tr}\,tt^{\dagger}(1-tt^{\dagger})
=P_{0}\sum_{n=1}^{N}T_{n}(1-T_{n}),\label{e2}
\end{equation}
with $P_{0}\equiv 2e|V|(2e^{2}/h)$. Equation (\ref{e2}) is the multi-channel
generalization of earlier single-channel formulas \cite{Khl87,Les89}. It is a
consequence of the Pauli principle that closed ($T_{n}=0$) as well as open
($T_{n}=1$) scattering channels do not fluctuate and therefore give no
contribution to the shot noise. In the case of a tunnel barrier, all
transmission eigenvalues are small ($T_{n}\ll 1$, for all $n$), so that the
quadratic terms in eq.\ (\ref{e2}) can be neglected. Then it follows from
comparison with eq.\ (\ref{Landauer}) that $P_{\rm N}=2e|V|G_{\rm
N}=2e|I|=P_{\rm Poisson}$. In contrast, for a quantum point contact $P_{\rm
N}\ll P_{\rm Poisson}$. Since on the plateaus of quantized conductance
all the $T_{n}$'s are either 0 or 1, the shot noise is expected to be
only observable at the steps between the plateaus \cite{Les89}. For a diffusive
conductor of length $L$ much longer than the elastic mean free path $l$, the
shot noise $P_{\rm N}=\frac{1}{3}P_{\rm Poisson}$ is one-third the Poisson
noise, as a consequence of noiseless open scattering channels
\cite{B&B92,Nag92}.

The analogue of eq.\ (\ref{e2}) for the shot-noise power of an NS junction is
\cite{Jon94}
\begin{equation}
P_{\rm NS}=4P_{0} {\rm Tr}\,
s_{\rm he}^{\vphantom{\dagger}}s_{\rm he}^{\dagger}
(1-s_{\rm he}^{\vphantom{\dagger}}s_{\rm he}^{\dagger})=P_{0}\sum_{n=1}^{N}
\frac{16T_{n}^{2}(1 - T_{n})}{(2-T_{n})^{4}},\label{e17}
\end{equation}
where we have used eq.\ (\ref{she}) (with $\varepsilon=0$) to relate the
scattering matrix $s_{\rm he}$ for Andreev reflection to the transmission
eigenvalues $T_{n}$ of the normal region. This requires zero magnetic field. As
in the normal state, scattering channels which have $T_{n}=0$ or $T_{n}=1$ do
not contribute to the shot noise. However, the way in which partially
transmitting channels contribute is entirely different from the normal state
result (\ref{e2}).

Consider first an NS junction without disorder, but with an arbitrary
transmission probability $\Gamma$ per mode of the interface. In the normal
state, eq.\ (\ref{e2}) yields $P_{\rm N}=(1-\Gamma)P_{\rm Poisson}$, implying
full Poisson noise for a high tunnel barrier ($\Gamma\ll 1$). For the NS
junction we find from eq.\ (\ref{e17})
\begin{equation}
P_{\rm NS}=P_{0}N\frac{16\Gamma^{2}(1-\Gamma)}{(2-\Gamma)^{4}}=
\frac{8(1-\Gamma)}{(2-\Gamma)^{2}}P_{\rm Poisson},\label{e18}
\end{equation}
where in the second equality we have used eq.\ (\ref{keyzero}). This agrees
with results obtained by Khlus \cite{Khl87}, and by Muzykantski\u{\i} and
Khmel'nitski\u{\i} \cite{Muz94}, using different methods. If
$\Gamma<2(\sqrt{2}-1)\approx 0.83$, one observes a shot noise above the Poisson
noise. For $\Gamma\ll 1$ one has
\begin{equation}
P_{\rm NS}=4e|I|=2 P_{\rm Poisson},\label{e19}
\end{equation}
which is a doubling of the shot-noise power divided by the current with respect
to the normal-state result. This can be interpreted as uncorrelated current
pulses of $2e$-charged particles.

Consider next an NS junction with a disordered normal region, but with an ideal
interface ($\Gamma=1$). We may then apply the formula (\ref{avf}) for the
average of a linear statistic on the transmission eigenvalues to eqs.\
(\ref{keyzero}) and (\ref{e17}). The result is
\begin{equation}
\frac{\left\langle P_{\rm NS}\right\rangle}
{\left\langle G_{\rm NS}\right\rangle}=
\frac{2}{3}\,\frac{P_0}{2e^{2}/h}\;\Rightarrow\;\left\langle P_{\rm
NS}\right\rangle=
{\textstyle\frac{4}{3}}e|I|={\textstyle\frac{2}{3}}P_{\rm Poisson}.
\label{e20a}
\end{equation}
Equation (\ref{e20a}) is twice the result in the normal state, but still
smaller than the Poisson noise. Corrections to (\ref{e20a}) are of lower order
in $N$ and due to quantum-interference effects \cite{Jon92}.

\begin{figure}[tb]
\hspace*{\fill}
\psfig{figure=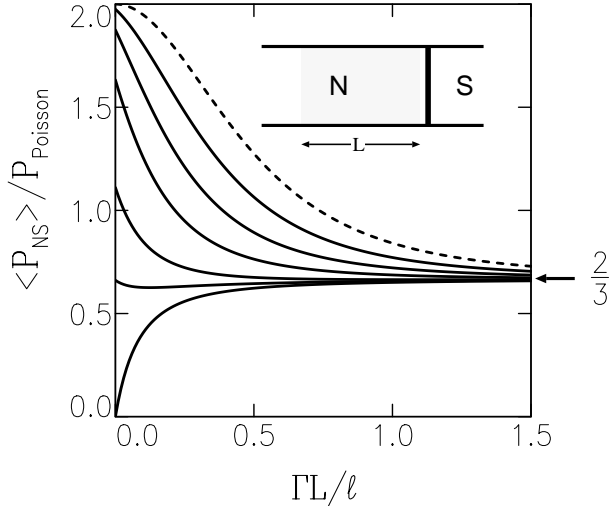,width=8cm}
\hspace*{\fill}
\medskip\caption[]{The shot-noise power of an NS junction (in units of $P_{\rm
Poisson}\equiv 2e|I|$) as a function of the length $L$ (in units of
$l/\Gamma$), for barrier transparencies  $\Gamma= 1,0.9,0.8,0.6,0.4,0.2$
from bottom to top. The dashed curve gives the limiting result for $\Gamma\ll
1$. For $L=0$ the noise power varies as a function of $\Gamma$ according to
eq.\ (\protect\ref{e18}), between doubled shot noise ($\langle P_{\rm
NS}\rangle=4e|I|$) for a high barrier ($\Gamma\ll 1$) and zero in the absence
of a barrier ($\Gamma=1$). For $L\rightarrow\infty$ the noise power approaches
the limiting value $\langle P_{\rm NS}\rangle=\frac{4}{3}e|I|$ for
each $\Gamma$. {\em (From ref.\ \protect\cite{Jon94}.)}
\label{shotnoise}}
\end{figure}

Finally, consider an NS junction which contains a disordered normal region
(length $L$, mean free path $l$) as well as a non-ideal interface. The scaling
theory of section 5.2 has been applied to this problem in ref.\ \cite{Jon94}.
Results are shown in fig.\ \ref{shotnoise}, where
$\langle P_{\rm NS}\rangle/P_{\rm Poisson}$ is plotted against $\Gamma L/l$ for
various $\Gamma$. Note the crossover from the ballistic result (\ref{e18}) to
the diffusive result (\ref{e20a}). For a high barrier ($\Gamma \ll 1$), the
shot noise decreases from twice the Poisson noise to two-thirds the Poisson
noise as the amount of disorder increases.

\section{Conclusion}

We have reviewed a scattering approach to phase-coherent transport accross the
interface between a normal metal and a superconductor. For the
reflectionless-tunneling phenomenon, the complete equivalence has been
demonstrated to the non-equilibrium Green's function approach. (The other
effects we discussed have so far mainly been treated in the scattering
approach.) Although mathematically equivalent, the physical picture offered by
the two approaches is quite different. We chose to focus on the scattering
approach because it makes direct contact with the quantum interference effects
studied extensively in the normal state. The same techniques used for weak
localization and universal conductance fluctuations in normal conductors could
be used to study the modifications by Andreev reflection in an NS junction.

In the limit of zero voltage, zero temperature, and zero magnetic field, the
transport properties of the NS junction are determined entirely by the
transmission eigenvalues $T_{n}$ of the normal region. A scaling theory for the
distribution of the $T_{n}$'s then allows one to obtain analytical results for
the mean and variance of any observable of the form $A=\sum_{n}a(T_{n})$. The
conductance is of this form, as well as the shot-noise power. The only
difference with the normal state is the functional form of $a(T)$ (polynomial
in the normal state, rational function for an NS junction), so that the general
results of the scaling theory [valid for any function $a(T)$] can be applied at
once. At finite $V$, $T$, or $B$, one needs the entire scattering matrix of the
normal region, not just the transmission eigenvalues. This poses no difficulty
for a numerical calculation, as we have shown in several examples. However,
analytical progress using the scattering approach becomes cumbersome, and a
diagrammatic Green's function calculation is more efficient.

{\em Note added February 1995:} The theory of section 4 has been
extended to non-zero voltage and magnetic field by P. W. Brouwer and
the author (submitted to Phys.\ Rev.\ B). The results are $\delta
G_{\rm NS}(V=0,B\neq 0)=\frac{1}{3}e^{2}/h$, $\delta G_{\rm NS}(V\neq
0,B=0)=\frac{2}{3}e^{2}/h$, $\delta G_{\rm NS}(V\neq 0,B\neq 0)=0$. The
disagreement with the numerical simulations discussed in section 4 is
due to an insufficiently large system size.

\section*{Acknowledgments}
It is a pleasure to acknowledge my collaborators in this research: R. A.
Jalabert, M. J. M. de Jong, I. K. Marmorkos, J. A. Melsen, and B. Rejaei.
Financial support was provided by the ``Ne\-der\-land\-se or\-ga\-ni\-sa\-tie
voor We\-ten\-schap\-pe\-lijk On\-der\-zoek'' (NWO), by the ``Stich\-ting voor
Fun\-da\-men\-teel On\-der\-zoek der Ma\-te\-rie'' (FOM), and by the European
Community. I have greatly benefitted from the insights of Yu.\ V. Nazarov.
Permission to reproduce the experimental figs.\ \ref{Lenssen}, \ref{Kastalsky},
and \ref{Pothier} was kindly given by K.-M. H. Lenssen, A. W. Kleinsasser, and
H. Pothier, respectively.


\begin{references}
\bibitem{And64} A. F. Andreev, Zh.\ Eksp.\ Teor.\ Fiz.\ {\bf 46}, 1823 (1964)
[Sov.\ Phys.\ JETP {\bf 19}, 1228 (1964)].
\bibitem{Eer91} C. W. J. Beenakker and H. van Houten, Solid State Phys.\ {\bf
44}, 1 (1991).
\bibitem{Lam93a} C. J. Lambert, J. Phys.\ Condens.\ Matter {\bf 5}, 707 (1993).
\bibitem{Lam93b} C. J. Lambert, V. C. Hui, and S. J. Robinson, J. Phys.\
Condens.\ Matter {\bf 5}, 4187 (1993).
\bibitem{Pet93} V. T. Petrashov, V. N. Antonov, P. Delsing, and T. Claeson,
Phys.\ Rev.\ Lett.\ {\bf 70}, 347 (1993).
\bibitem{Kla94} T. M. Klapwijk, Physica B {\bf 197}, 481 (1994).
\bibitem{Bee91} C. W. J. Beenakker, Phys.\ Rev.\ Lett.\ {\bf 67}, 3836 (1991);
{\bf 68}, 1442(E) (1992).
\bibitem{Shima} C. W. J. Beenakker, in: {\em Transport Phenomena in Mesoscopic
Systems}, ed.\ by H. Fukuyama and T. Ando (Springer, Berlin, 1992).
\bibitem{Pip71} The contribution of scattering inside the superconductor to the
resistance of an NS junction has been studied extensively, see for example: A.
B. Pippard, J. G. Shepherd, and D. A. Tindall, Proc.\ R. Soc.\ London A {\bf
324}, 17 (1971); A. Schmid and G. Sch\"{o}n, J. Low Temp.\ Phys.\ {\bf 20}, 207
(1975); T. Y. Hsiang and J. Clarke, Phys.\ Rev.\ B {\bf 21}, 945 (1980).
\bibitem{deG66} P. G. de Gennes, {\em Superconductivity of Metals and Alloys}
(Benjamin, New York, 1966).
\bibitem{Lik79} K. K. Likharev, Rev.\ Mod.\ Phys.\ {\bf 51}, 101 (1979).
\bibitem{Blo82} G. E. Blonder, M. Tinkham, and T. M. Klapwijk, Phys.\ Rev.\ B
{\bf 25}, 4515 (1982).
\bibitem{Lam91} C. J. Lambert, J. Phys.\ Condens.\ Matter {\bf 3}, 6579 (1991).
\bibitem{Tak92a} Y. Takane and H. Ebisawa, J. Phys.\ Soc.\ Jpn.\ {\bf 61}, 1685
(1992).
\bibitem{Bee92} C. W. J. Beenakker, Phys.\ Rev.\ B {\bf 46}, 12841 (1992).
\bibitem{She84} A. L. Shelankov, Fiz.\ Tverd.\ Tela {\bf 26}, 1615
(1984) [Sov.\ Phys.\ Solid State {\bf 26}, 981 (1984)].
\bibitem{Zai84} A. V. Za\u{\i}tsev, Zh.\ Eksp.\ Teor.\ Fiz.\ {\bf 86}, 1742
(1984) [Sov.\ Phys.\ JETP {\bf 59}, 1015 (1984)].
\bibitem{Hou91} H. van Houten and C. W. J. Beenakker, Physica B {\bf
175}, 187 (1991).
\bibitem{Zai80} A. V. Za\u{\i}tsev, Zh.\ Eksp.\ Teor.\ Fiz.\ {\bf 78}, 221
(1980); {\bf 79}, 2016(E) (1980) [Sov.\ Phys.\ JETP {\bf 51}, 111 (1980); {\bf
52}, 1018(E) (1980)].
\bibitem{But90} M. B\"{u}ttiker, Phys.\ Rev.\ B {\bf 41}, 7906 (1990).
\bibitem{But88} M. B\"{u}ttiker, IBM J. Res.\ Dev.\ {\bf 32}, 63 (1988).
\bibitem{Gla89} L. I. Glazman and K. A. Matveev, Pis'ma Zh.\ Eksp.\ Teor.\
Fiz.\ {\bf 49}, 570 (1989) [JETP Lett.\ {\bf 49}, 659 (1989)].
\bibitem{SQUID91} C. W. J. Beenakker and H. van Houten, in: {\em
Single-Electron Tunneling and Mesoscopic Devices}, ed.\ by H. Koch and H.
L\"{u}bbig (Springer, Berlin, 1992).
\bibitem{Khl93} V. A. Khlus, A. V. Dyomin, and A. L. Zazunov, Physica C {\bf
214}, 413 (1993).
\bibitem{Dev90} I. A. Devyatov and M. Yu.\ Kupriyanov, Pis'ma Zh.\ Eksp.\
Teor.\ Fiz.\ {\bf 52}, 929 (1990) [JETP Lett.\ {\bf 52}, 311 (1990)].
\bibitem{Hek93a} F. W. J. Hekking, L. I. Glazman, K. A. Matveev, and R. I.
Shekhter, Phys.\ Rev.\ Lett.\ {\bf 70}, 4138 (1993).
\bibitem{Dor84} O. N. Dorokhov, Solid State Comm.\ {\bf 51}, 381 (1984).
\bibitem{Imr86} Y. Imry, Europhys.\ Lett.\ {\bf 1}, 249 (1986).
\bibitem{Pen92} J. B. Pendry, A. MacKinnon, and P. J. Roberts, Proc.\ R.\ Soc.\
London A {\bf 437}, 67 (1992).
\bibitem{Naz94} Yu.\ V. Nazarov (to be published).
\bibitem{And66} A. F. Andreev, Zh.\ Eksp.\ Teor.\ Fiz.\ {\bf 51}, 1510 (1966)
[Sov.\ Phys.\ JETP {\bf 24}, 1019 (1967)].
\bibitem{Art79} S. N. Artemenko, A. F. Volkov, and A. V. Za\u{\i}tsev, Solid
State Comm.\ {\bf 30}, 771 (1979).
\bibitem{Tak92b} Y. Takane and H. Ebisawa, J. Phys.\ Soc.\ Jpn.\ {\bf 61}, 2858
(1992).
\bibitem{Sto91} A. D. Stone, P. A. Mello, K. A. Muttalib, and J.-L. Pichard,
in: {\em Mesoscopic Phenomena in Solids}, ed. by B. L. Al'tshuler, P. A. Lee,
and R. A. Webb (North-Holland, Amsterdam, 1991).
\bibitem{Mel91} P. A. Mello and A. D. Stone, Phys.\ Rev.\ B {\bf 44}, 3559
(1991).
\bibitem{Bee94} C. W. J. Beenakker, Phys.\ Rev.\ B {\bf 49}, 2205 (1994).
\bibitem{Mac94} A. M. S. Mac\^{e}do and J. T. Chalker, Phys.\ Rev.\ B {\bf 49},
4695 (1994).
\bibitem{Tak94} Y. Takane and H. Otani [J. Phys.\ Soc.\ Jpn.\ (to be
published)] find $\delta G_{\rm NS}=\frac{4}{3}\,e^{2}/h$, in slight
disagreement with eq.\ (\protect\ref{weaklocaNSN}).
\bibitem{Mar93} I. K. Marmorkos, C. W. J. Beenakker, and R. A. Jalabert, Phys.\
Rev.\ B {\bf 48}, 2811 (1993).
\bibitem{Len94} K.-M. H. Lenssen, P. C. A. Jeekel, C. J. P. M. Harmans, J. E.
Mooij, M. R. Leys, J. H. Wolter, and M. C. Holland, in: {\em Coulomb and
Interference Effects in Small Electronic Structures}, ed.\ by D. C. Glattli and
M. Sanquer (Editions Fronti\`{e}res, to be published).
\bibitem{Kas91} A. Kastalsky, A. W. Kleinsasser, L. H. Greene, R. Bhat,
F. P. Milliken, and J. P. Harbison, Phys.\ Rev.\ Lett.\ {\bf 67}, 3026 (1991).
\bibitem{Ngu92} C. Nguyen, H. Kroemer, and E. L.  Hu, Phys.\ Rev.\ Lett.\ {\bf
69}, 2847 (1992).
\bibitem{Man92} R. G. Mani, L. Ghenim, and T. N. Theis, Phys.\ Rev.\ B {\bf
45}, 12098 (1992).
\bibitem{Agr92} N. Agra\"{\i}t, J. G. Rodrigo, and S. Vieira, Phys.\ Rev.\ B
{\bf 46}, 5814 (1992).
\bibitem{Xio93} P. Xiong, G. Xiao, and R. B. Laibowitz, Phys.\ Rev.\ Lett.\
{\bf 71}, 1907 (1993).
\bibitem{Len94b} K.-M. H. Lenssen, L. A. Westerling, P. C. A. Jeekel, C. J. P.
M. Harmans, J. E. Mooij, M. R. Leys, W. van der Vleuten, J. H. Wolter, and S.
P. Beaumont, Physica B {\bf 194--196}, 2413 (1994).
\bibitem{Bak94} S. J. M. Bakker, E. van der Drift, T. M. Klapwijk, H. M.
Jaeger, and S. Radelaar, Phys.\ Rev.\ B {\bf 49}, 13275 (1994).
\bibitem{Mag94} P. H. C. Magn\'{e}e, N. van der Post, P. H. M. Kooistra, B. J.
van Wees, and T. M. Klapwijk, Phys.\ Rev.\ B (to be published).
\bibitem{Tak93a} Y. Takane and H. Ebisawa, J. Phys.\ Soc.\ Jpn.\ {\bf 62}, 1844
(1993).
\bibitem{Wee92} B. J. van Wees, P. de Vries, P. Magn\'{e}e, and T. M. Klapwijk,
Phys.\ Rev.\ Lett.\ {\bf 69}, 510 (1992).
\bibitem{Tak92c} Y. Takane and H. Ebisawa, J. Phys.\ Soc.\ Jpn.\ {\bf 61}, 3466
(1992).
\bibitem{Vol93} A. F. Volkov, A. V. Za\u{\i}tsev, and T. M. Klapwijk, Physica C
{\bf 210}, 21 (1993).
\bibitem{Hek93} F. W. J. Hekking and Yu.\ V. Nazarov, Phys.\ Rev.\ Lett.\ {\bf
71}, 1625 (1993); Phys.\ Rev. B {\bf 49}, 6847 (1994).
\bibitem{Bee94b} C. W. J. Beenakker, B. Rejaei, and J. A. Melsen, Phys.\ Rev.\
Lett.\ {\bf 72}, 2470 (1994).
\bibitem{Zai90} A. V. Za\u{\i}tsev, Pis'ma Zh.\ Eksp.\ Teor.\ Fiz.\ {\bf 51},
35 (1990) [JETP Lett.\ {\bf 51}, 41 (1990)]; Physica C {\bf 185--189}, 2539
(1991).
\bibitem{Vol92a} A. F. Volkov and T. M. Klapwijk, Phys.\ Lett.\ A {\bf 168},
217 (1992).
\bibitem{Vol92b} A. F. Volkov, Pis'ma Zh.\ Eksp.\ Teor.\ Fiz.\ {\bf 55}, 713
(1992) [JETP Lett.\ {\bf 55}, 746 (1992)]; Phys.\ Lett.\ A {\bf 174}, 144
(1993).
\bibitem{She80} A. L. Shelankov, Pis'ma Zh.\ Eksp.\ Teor.\ Fiz.\ {\bf 32}, 122
(1980) [JETP Lett.\ {\bf 32}, 111 (1980)].
\bibitem{Mel89} P. A. Mello and J.-L. Pichard, Phys.\ Rev.\ B {\bf 40}, 5276
(1989).
\bibitem{Dor82} O. N. Dorokhov, Pis'ma Zh.\ Eksp.\ Teor.\ Fiz.\ {\bf 36}, 259
(1982) [JETP Lett.\ {\bf 36}, 318 (1982)].
\bibitem{Mel88} P. A. Mello, P. Pereyra, and N. Kumar, Ann.\ Phys.\ {\bf 181},
290 (1988).
\bibitem{Mel94} J. A. Melsen and C. W. J. Beenakker, Physica B (to be
published).
\bibitem{Hesl94} D. R. Heslinga, S. E. Shafranjuk, H. van Kempen, and T. M.
Klapwijk, Phys.\ Rev.\ B {\bf 49},
 10484 (1994).
\bibitem{Vol94} A. F. Volkov, Phys.\ Lett.\ A {\bf 187}, 404 (1994).
\bibitem{Naz94b} Yu.\ V. Nazarov (to be published).
\bibitem{Lar77} A. I. Larkin and Yu.\ N. Ovchinnikov, Zh.\ Eksp.\ Teor.\ Fiz.\
{\bf 68}, 1915 (1975); {\bf 73}, 299 (1977) [Sov.\ Phys.\ JETP {\bf 41}, 960
(1975); {\bf 46}, 155 (1977)].
\bibitem{Kup88} M. Yu.\ Kupriyanov and V. F. Lukichev, Zh.\ Eksp.\ Teor.\ Fiz.\
{\bf 94}, 139 (1988) [Sov.\ Phys.\ JETP {\bf 67}, 1163 (1988)].
\bibitem{Naz94c} Yu.\ V. Nazarov, contribution at the NATO Adv.\ Res.\ Workshop
on ``Mesoscopic Superconductivity'' (Karlsruhe, May 1994).
\bibitem{Pot94} H. Pothier, S. Gu\'{e}ron, D. Esteve, and M. H. Devoret (to be
published).
\bibitem{Spi82} B. Z. Spivak and D. E. Khmel'nitski\u{\i}, Pis'ma Zh.\ Eksp.\
Teor.\ Fiz.\ {\bf 35}, 334 (1982) [JETP Lett.\ {\bf 35}, 412 (1982)].
\bibitem{Alt87} B. L. Al'tshuler and B. Z. Spivak, Zh.\ Eksp.\ Teor.\ Fiz.\
{\bf 92}, 609 (1987) [Sov.\ Phys.\ JETP {\bf 65}, 343 (1987)].
\bibitem{Nak91} H. Nakano and H. Takayanagi, Solid State Comm.\ {\bf 80}, 997
(1991).
\bibitem{Tak92} S. Takagi, Solid State Comm.\ {\bf 81}, 579 (1992).
\bibitem{Lam93} C. J. Lambert, J. Phys.\ Condens.\ Matter {\bf 5}, 707 (1993).
\bibitem{Hui93} V. C. Hui and C. J. Lambert, Europhys.\ Lett.\ {\bf 23}, 203
(1993).
\bibitem{Bee93} C. W. J. Beenakker, Phys.\ Rev.\ Lett.\ {\bf 70}, 1155 (1993);
Phys.\ Rev.\ B {\bf 47}, 15763 (1993).
\bibitem{Bee93b} C. W. J. Beenakker and B. Rejaei, Phys.\ Rev.\ Lett.\ {\bf
71}, 3689 (1993); Phys.\ Rev.\ B {\bf 49}, 7499 (1994).
\bibitem{Cha94} J. T. Chalker and A. M. S. Mac\^{e}do, Phys.\ Rev.\ Lett.\ {\bf
71}, 3693 (1993).
\bibitem{Tak91} Y. Takane and H. Ebisawa, J. Phys.\ Soc.\ Jpn.\ {\bf 60}, 3130
(1991).
\bibitem{Bru94} J. Bruun, V. C. Hui, and C. J. Lambert, Phys.\ Rev.\ B {\bf
49}, 4010 (1994).
\bibitem{Jon94} M. J. M. de Jong and C. W. J. Beenakker, Phys.\ Rev.\ B (to be
published).
\bibitem{But90b} M. B\"{u}ttiker, Phys.\ Rev.\ Lett.\ {\bf 65}, 2901 (1990);
Phys. Rev. B {\bf 46}, 12485 (1992).
\bibitem{Khl87} V. A. Khlus, Zh.\ Eksp.\ Teor.\ Fiz.\ {\bf 93}, 2179 (1987)
[Sov.\ Phys.\ JETP {\bf 66}, 1243 (1987)].
\bibitem{Les89} G. B. Lesovik, Pis'ma Zh.\ Eksp.\ Teor.\ Fiz.\ {\bf 49}, 513
(1989) [JETP Lett.\ {\bf 49}, 592 (1989)].
\bibitem{B&B92} C. W. J. Beenakker and M. B\"{u}ttiker, Phys.\ Rev.\ B {\bf
46}, 1889 (1992).
\bibitem{Nag92} K. E. Nagaev, Phys.\ Lett.\ A {\bf 169}, 103 (1992).
\bibitem{Muz94} B. A. Muzykantski\u{\i} and D. E. Khmel'nitski\u{\i} (to be
published).
\bibitem{Jon92} M. J. M. de Jong and C. W. J. Beenakker, Phys.\ Rev.\ B {\bf
46}, 13400 (1992).

\end{references}
\end{document}